\documentclass{jfm}
\usepackage{xcolor}
\usepackage{amsmath}
\usepackage{soul}
\usepackage{float}
\usepackage{subcaption}
\usepackage[normalem]{ulem}
\usepackage{amssymb}
\usepackage{url}

\DeclareFontFamily{OT1}{pzc}{}
\DeclareFontShape{OT1}{pzc}{m}{it}{<-> s * [1.10] pzcmi7t}{}
\DeclareMathAlphabet{\mathpzc}{OT1}{pzc}{m}{it}

\begin{document}

\newtheorem{lemma}{Lemma}
\newtheorem{corollary}{Corollary}
\newcommand{\setbackgroundcolour}{\pagecolor[rgb]{0.19,0.19,.19}}  
\newcommand{\settextcolour}{\color[rgb]{0.77,0.77,0.77}}    
\newcommand{\invertbackgroundtext}{\setbackgroundcolour\settextcolour}

\newcommand{\Bo}{B\kern-0.04em o\,}
\newcommand{\Bos}{B\kern-0.04em o}
\newcommand{\We}{W\kern-0.03em e\,}
\newcommand{\Wes}{W\kern-0.03em e}
\newcommand{\Fr}{F\kern-0.04em r\,}
\newcommand{\Frs}{F\kern-0.04em r}
\newcommand{\Dr}{D\kern-0.04em r\,}
\newcommand{\Drs}{D\kern-0.04em r}
\newcommand{\Oh}{O\kern-0.04em h\,}

\shorttitle{Inertio-capillary droplet rebound} 
\shortauthor{Alventosa, Cimpeanu, and Harris} 

\title{Inertio-capillary rebound of a droplet impacting a fluid bath}

 \author
   {
   Luke F.L. Alventosa\aff{1},
   Radu Cimpeanu\aff{2,3,4},
   \and 
   Daniel M. Harris\aff{1}
   }

 \affiliation
 {
 \aff{1}
 School of Engineering, Brown University, Providence, Rhode Island 02912, USA
 \aff{2}
 Mathematics Institute, University of Warwick, Coventry CV4 7AL, United Kingdom
 \aff{3}
 Mathematical Institute, University of Oxford, Oxford OX2 6GG, United Kingdom
 \aff{4}
Department of Mathematics, Imperial College London, London SW7 2AZ, United Kingdom
}

\maketitle
\begin{abstract}
The rebound of droplets impacting a deep fluid bath is studied both experimentally and theoretically. Millimetric drops are generated using a piezoelectric droplet-on-demand generator and normally impact a bath of the same fluid.  Measurements of the droplet trajectory and other rebound metrics are compared directly to the predictions of a linear quasi-potential model, as well as fully resolved direct numerical simulations (DNS) of the unsteady Navier-Stokes equations. Both models resolve the time-dependent bath and droplet shapes in addition to the droplet trajectory. In the quasi-potential model, the droplet and bath shape are decomposed using orthogonal function decompositions leading to two sets of coupled damped linear harmonic oscillator equations solved using an implicit numerical method. The underdamped dynamics of the drop are directly coupled to the response of the bath through a single-point kinematic match condition which we demonstrate to be an effective and efficient model in our parameter regime of interest. Starting from the inertio-capillary limit in which both gravitational and viscous effects are negligible, increases in gravity or viscosity lead to a decrease in the coefficient of restitution and an increase in the contact time. The inertio-capillary limit defines an upper bound on the possible coefficient of restitution for droplet-bath impact, depending only on the Weber number. The quasi-potential model is able to rationalize historical experimental measurements for the coefficient of restitution, first presented by \cite{jayaratne1964coalescence}.

\end{abstract}
\section{Introduction}

Droplet impacts occur frequently in both natural and industrial settings. Rain drops impacting on leaves have been shown to be a primary mechanism for pathogen transport among plants \citep{kim2019vortex} and birds with superhydrophobic feathers stay warmer in a cold rain due to a reduced droplet contact time \citep{shiri2017heat}. Spray cooling devices have attracted the attention of researchers due to the large heat transfer rates and high uniformity of heat transfer \citep{kim2007spray}. Wet scrubbing of exhaust gases relies on the inertial impaction of small particles and aerosols on the surface of freely falling droplets \citep{park2005wet}. Various other drop impact phenomena, such as splashing, were experimentally documented by Worthington at the start of the 20$^{th}$ century \citep{worthington1908study}. More recently, droplets bouncing repeatedly on a vertically oscillated bath have received considerable interest as a macroscopic pilot-wave system capable of reproducing some behaviors reminiscent of quantum particles \citep{couder2005walking,bush2020hydrodynamic}.  For instance, bouncing droplets confined to submerged cavities can exhibit wave-like statistical behavior analogous to electrons in quantum corrals \citep{harris2013wavelike,saenz2018statistical}.  Droplet impact onto solid surfaces is also an extremely well-studied field \citep{yarin2006drop,josserand2016drop}, with the combination of high quality experiments and direct numerical simulation (DNS) leading to a deep understanding of the multi-scale dynamics. 
\\
\begin{figure}
    \centering
    \includegraphics[width=1\textwidth]{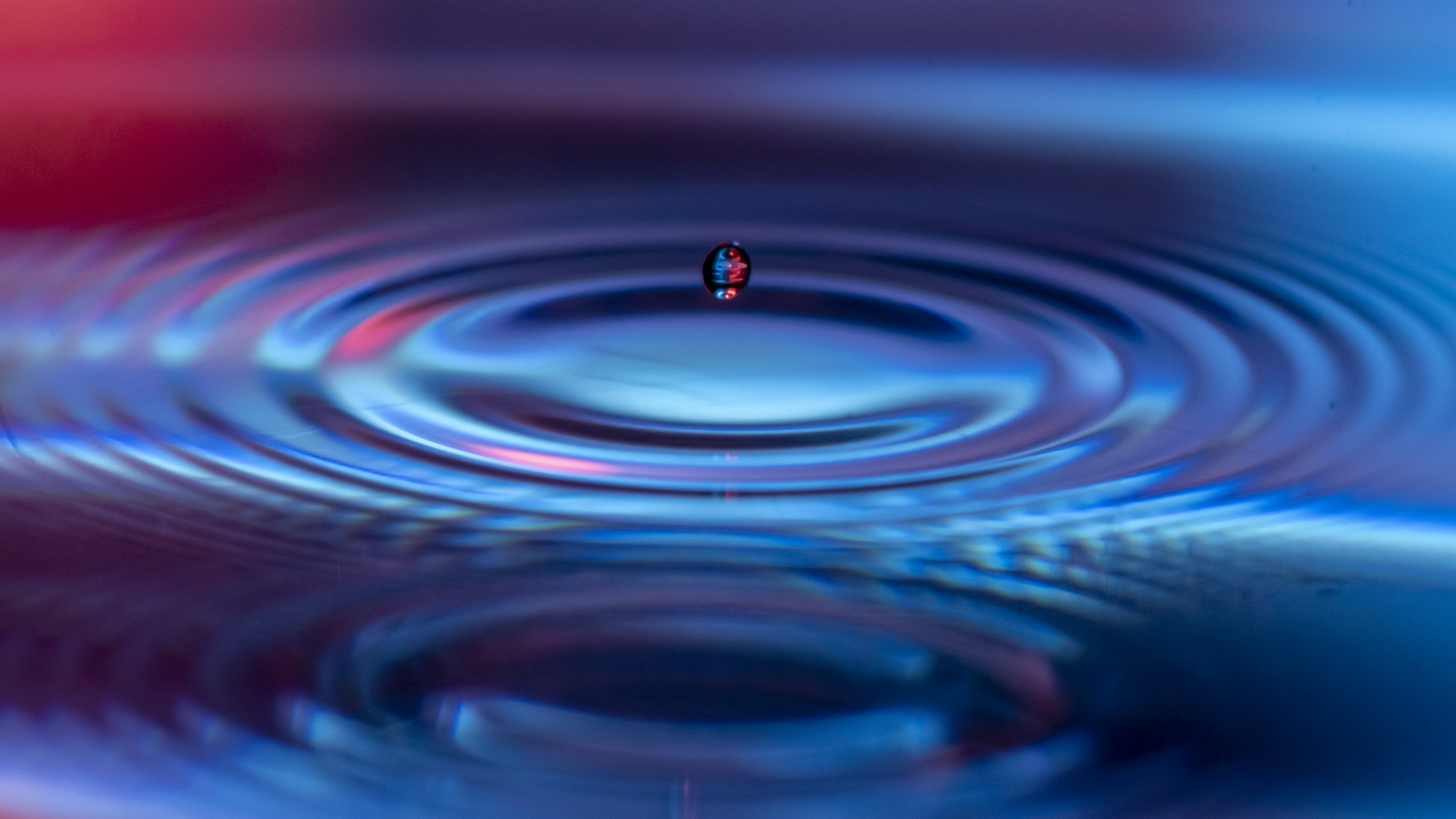}
    \caption{A small water droplet ($R\approx0.4$ mm) rebounds from a bath of the same fluid.}
    \label{fig:art}
\end{figure}
\hspace*{1em} The problem of droplet coalescence onto a bath of the same fluid has also been studied extensively over the last century and a half, beginning with \cite{rayleigh1879capillary} and \cite{thomson1886v}.  These early works included sketches of drop-interface coalescence, as well as a detailed description of the vortices that are formed in the fluid bath. Preceding coalescence, the thin gas film that forms between the two interfaces drains until van der Waals forces act to initiate coalescence.  Coalescence of a drop into a bath occurs when the film is on the order of $100$ nm thick \citep{couder2005bouncing,yarin2006drop,de2012dynamics,kavehpour2015coalescence}. The development of accessible high-speed photography and high performance computing has ushered in a rapid expansion in quantity and quality of data on these free surface problems. \cite{thoroddsen2000coalescence} used a high-precision and high-speed visualization setup to quantify the coalescence time of droplets on an air-liquid interface. \cite{tang2019bouncing} studied the dynamics of the gas layer on a liquid bath whose depth was similar to that of the droplet radius. The rich class of outcomes and dynamics that arise from such a simple interaction between droplet, surrounding gas, and interface proves that these fundamental problems merit considerable attention. \\
\hspace*{1em} During contact, the combined effects of inertia, surface tension, gravity, and viscosity govern the hydrodynamic interaction between the droplet and the interface, and the complex balance of forces within this regime creates the variety of distinct phenomena. The Weber number $\We$ $= \rho V_0^2 R /\sigma$, the Bond number $\Bo$ $= \rho R^2 g/\sigma$, and the Ohnesorge number $\Oh$ $= \mu/\sqrt{\sigma R \rho}$ are often used to describe these capillary-scale dynamics. In this work, $R$ represents the undeformed droplet radius, $\rho$, $\sigma$, $\mu$ are the density, surface tension, and viscosity of the fluid in both the droplet and bath, $V_0$ is the impact velocity of the droplet, and $g$ is the gravitational acceleration.  The present work focuses on the inertio-capillary regime, where fluid inertia and surface tension dominate viscous and gravitational effects (specifically, $\Oh\ll1$ and $\Bo\ll1$).  During impact, a thin gas film develops between the free interface and surface of the droplet. The drainage of this thin film plays a crucial role in determining the fate of the droplet: specifically whether it rebounds from or coalesces with the underlying bath \citep{de2012dynamics}.  At sufficiently low $\We$, the droplet and the interface never come into physical contact and remain separated by a stable air film. The droplet then levitates on this thin film and can eventually rebound due to the relaxation of the bath and droplet interface. Droplets bouncing on a free interface were first documented by Reynolds in 1881, when he noted that drops can ``float'' on a bath of the same liquid if the impact velocity is sufficiently small \citep{reynolds1881drops}. \\ 
\hspace*{1em} In cases of droplet-bath impact, as well as droplet-droplet impact, there exists a parameter regime where droplets bounce completely (Figure \ref{fig:art}).  The bouncing-coalescence threshold is often characterized by a critical $\We$ that depends sensitively on all parameters in the problem \citep{tang2019bouncing}.  Droplet bouncing on an undisturbed interface at variable impacting angles was first studied in detail by \cite{jayaratne1964coalescence} experimentally. They were able to determine a relationship between the drop radius, impact speed, and impact angle at which the bouncing-coalescence threshold occurs between uncharged drops. 
Building on the work of \cite{gopinath2001dynamics}, \cite{bach2004coalescence} studied droplet impact of small ($ R \leq 50 \text{ } \mu\mathrm{m}$) aerosol droplets impacting a fluid bath. They developed a rarefied gas model to describe the dynamics of the gas layer separating the droplet and bath, and used an inviscid potential flow model to describe the transfer of energy from the droplet to the bath during impact. The authors determined that the criterion for drop bouncing is more sensitive to gas mean-free path and gas viscosity than to the Weber number itself. \cite{zou2011experimental} investigated water droplets bouncing on an air-water interface, and examined the role of bath depth in bounce-back behavior. They determined that the contact time was independent of the impact velocity for a large range of Bond numbers. \cite{wu2020small} used a drop-on-demand generator to study the bouncing of water droplets, developed a model for the maximum penetration depth, and compared it to their experimental study. They varied the droplet diameter, and found that the maximum rebound height decreased with increasing diameter.  An experimental work utilizing three different fluids was completed by \cite{zhao2011transition}. They chose water, 1-propanol, and ethanol as the working fluids and found good agreement in measured contact times to \cite{jayaratne1964coalescence}. Also, they determined that the contact time of the droplet was relatively independent of the impact velocity, similar to that found by \cite{richard2002contact} for a droplet impacting a non-wetting, dry surface. In the variety of experimental work on this particular problem, the scaling for contact time $t_c$ of the droplet appears to be mostly independent of $\We$, except at very low $\We$ \citep{zhao2011transition,zou2011experimental,wu2020small}. Additionally, numerous papers report a saturation of translational energy recovery by the droplet at intermediate $\We$, as measured by the coefficient of restitution $\alpha$ \citep{jayaratne1964coalescence,bach2004coalescence,zhao2011transition,zou2011experimental,wu2020small}. These observations have not yet been fully explained nor their parametric dependencies clearly elucidated to the best of the current authors' knowledge. This motivates the development of a first principles model that can accurately and efficiently describe the dynamics of both the droplet and the fluid bath over the physically relevant parameter regime. \\
\hspace*{1em} The multi-scale hydrodynamics present in these impact problems creates significant challenges for numerical simulations, and all but eliminates analytical solutions to these problems. \cite{wagner1932stoss} proposed the first theoretical study of an object impacting on an inviscid, incompressible fluid, utilizing linearized free-surface kinematic and dynamic conditions to develop a theory that decomposed the fluid domain into two parts, one where the applied pressure is unknown but the interface shape is known, and vice versa. The so-called Wagner theory was extended to a solid of revolution by \cite{schmieden1953aufschlag} and eventually to three dimensions by \cite{scolan2001three}. These models assume that the working fluid is ideal, and thus any waves generated upon impact are not subject to viscous dissipation. \cite{dias2008theory} derived a theory to include the effects of weak damping in free surface problems, which appear as leading order corrections in the free surface boundary conditions. The inclusion of damping in this method provides a mechanism for the waves generated by impact to decay in time. The \cite{dias2008theory} theory is valid in the weakly viscous regime and represents a rigorous derivation of a linearized free surface model first proposed by \cite{lamb1895surface}. \\
\hspace*{1em} More recently, \cite{galeano2017non,galeano2019quasi,galeano2021capillary} applied the quasi-potential model of \cite{dias2008theory} to free surface impact problems and solves the problem of the unknown, time evolving contact region through the use of a so-called ``kinematic match''. In the kinematic match framework, the free surface shape within the region of contact is determined by the geometry of the problem, and the extent of this region can be computed with the use of a tangency boundary condition. The model in \cite{galeano2017non} worked well in determining the trajectory of the droplet in some cases, however neglected any deformations of the droplet. \cite{blanchette2016modeling,blanchette2017octahedra} modeled the impact of a droplet onto a still and oscillating bath, where a simplified version of the kinematic match concept was used by assuming that the shape and radial extent of the pressure distribution in the contact region are known {\it a priori}. Additionally, droplet deformations were modeled as a vertical spring or as an octahedral network of springs and masses. For still bath impacts, only very limited direct comparison to experimental measurements were made, with mixed success. \cite{molavcek2012quasi} developed a quasi-static model for a droplet impacting on a non-wetting rigid solid surface with fixed curvature. They compared this quasi-static model to a dynamic model that described the droplet-air interface using spherical harmonics derived from a balance of surface, kinetic, and potential energies and found good agreement between the two at low $\We$ numbers, as compared to experiments and the model of \cite{okumura2003water}. However, these models do not predict the energy transfer and time dependent waves on a fluid bath. \cite{terwagne2013role} wrote a linear mass-spring-damper model for a bouncing droplet on a vertically oscillated bath. Similarly, this model assumed that the bath surface was non-deformable. Additionally, \cite{molavcek2013drops} studied silicone oil droplets bouncing on a vertically oscillated bath and developed linear and logarithmic spring models to classify bouncing dynamics. While efficient to solve, these models require the input of free parameters determined from experimental data and thus cannot independently predict bouncing metrics such as the coefficient of restitution or contact time.  Other linear spring-type models have been proposed in the literature, but again such models generally rely on fitting parameters obtained from experimental data or direct numerical simulation \cite{sanjay2022drop}.  Direct numerical simulations of free surface impact problems have been completed in other recent works \citep{pan2007dynamics,he2015fluid,sharma2020energetics,fudge2021dipping}, and provide very good results, even in regimes presently inaccessible to experiments. From these simulations the droplet shape, trajectory, and waves, as well as the flow within the droplet and bath, can be captured and analyzed in detail. However, due to the high computational cost of these free surface flow problems, the vast parameter space encompassed by this problem renders large sweeps impractical for direct simulation, and leaves much to be understood about the overall dynamics over a more complete space. \\ 
\hspace*{1em} In this work, we develop an efficient model that accurately predicts the trajectory of the impacting droplet, the instantaneous droplet shape, and the transient waves generated on the bath interface by impact, without any adjustable parameters. First, we use the Navier-Stokes equations with linearized free surface conditions and include viscosity as leading order corrections to these boundary conditions, which holds in the limit of large $Re$.  We then derive a set of ordinary differential equations to describe the motion of the bath interface. The droplet shape is modeled by another set of ordinary differential equations that govern the weakly-damped oscillation of individual modes on the droplet interface that hold for small $Oh$.  Both the bath and droplet models are the result of linearizing about their undeformed states, and thus we anticipate best agreement when deformations are small.  The bath and drop models are coupled using a single-point kinematic match condition and evolved simultaneously in time.  We validate this model with new experimental data as well as direct numerical simulations.  We then apply the validated model over a wide range of parameters where the relative influence of the hydrodynamic, surface tension, and gravitational forces on the rebound behavior of the bouncing droplet will be elucidated.
\section{Experimental Methods}\label{sec.ExperimentalSetUP}
\subsection{Experimental setup}
\begin{figure}
    \centering
    \includegraphics[width=0.8\textwidth]{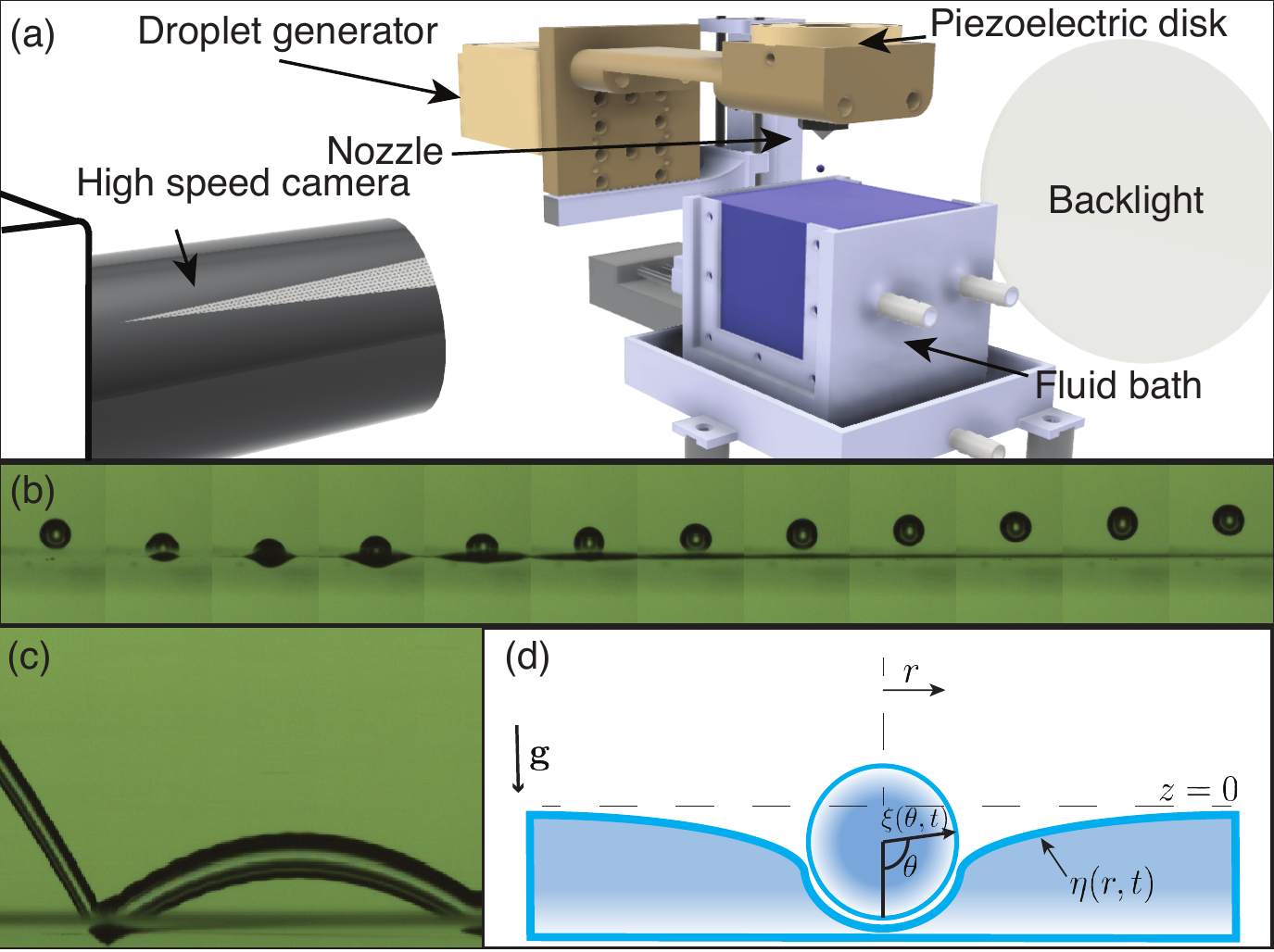}
    \caption{(a) A rendering of the experimental setup. (b) Experimental montage of impact of a deionized water droplet on a bath of the same fluid. Images are spaced 0.7 ms apart. (c) Spatiotemporal diagram of a deionized water droplet bouncing. The image is constructed by taking a single pixel wide stripe of the raw video footage, and plotting time along the $x$-axis. Panels (b) and (c) correspond to an impact of deionized water on a bath of the same fluid with $We=0.7$, $Bo = 0.017$, and $Oh = 0.006$. (d) Schematic of the problem.}
    \label{fig1:setup}
\end{figure}

A series of droplet impact experiments were conducted utilizing two working fluids: deionized water and silicone oil with viscosity of $5$ cSt. A drop-on-demand generator is used to reliably produce droplets with a maximum variation in the diameter of less than 1$\%$ \citep{ionkin2018note}. This device, along with a schematic of the experimental setup is shown in Figure \ref{fig1:setup}(a). The drop generator is entirely 3D printed, with the exception of a small piezoelectric disk, hardware, and connective tubing. 
The deformation of the piezoelectric disk due to an applied voltage pulse acts to expel fluid through a small nozzle. As the fluid exits the nozzle, the piezoelectric disk relaxes, initiating pinch off of the droplets. The droplets then fall under the action of gravity toward the bath. Two visualizations of droplet impact and rebound are shown in Figures \ref{fig1:setup}(b) and \ref{fig1:setup}(c). The drop generator is mounted on a 3D-printed translation stage, allowing for repeatable changes to the impacting velocity via height increases of the droplet generator. Directly underneath the drop generator is a 3D printed fluid bath. The bath is $70$ mm in width and length, and $50$ mm deep. The impact location was $25$ mm from the front wall of the bath. This impact location allowed for consistent focus above and below the free surface yet was still sufficiently far from the front panel that the waves created during impact do not have time to reflect and interact with the droplet during contact.  For the water experiments, the front and rear walls are constructed using polystyrene that has an equilibrium contact angle of $87.4^{\circ}$ \citep{ellison1954wettability}. Being close to $90^{\circ}$, this creates a negligibly small meniscus that allows for detailed photography of the impact from the side \citep{galeano2021capillary}. For the silicone oil experiments, we use a shorter bath window panel, constructed of extruded acrylic and a thin transparent plastic sheet. The bath was brim-filled to the height of the acrylic window panel, such that the contact line was pinned with angle held at approximately $90^{\circ}$. The drops are imaged using a high-speed camera (Phantom Miro LC 311) and illuminated by a Phlox LED-W back light. Video data is taken at 10,000 frames-per-second (fps) with an exposure time of $99.6 \text{ } \mathrm{ \mu s}$. \\

\begin{table}
\centering
\caption{Relevant parameters and their range of values in our experimental study.}
\begin{tabular}{lccc}
\textbf{Parameter}       
& \textbf{Symbol} 
& \textbf{Definition} 
& \textbf{Value}
\\
Impact speed          
& $V_0$          
& --
& $20-100\,$cm/s
\\
Droplet radius  
& $R$
& --
& $0.035\,$cm
\\
Density (water)         
& $\rho$        
& --
& $0.998\,$g/cm$^3$
\\
Surface tension (water) 
& $\sigma$        
& --                              
& $72.2 \,$dynes/cm
\\
Kinematic viscosity (water)
& $\nu$ 
& --
& $0.978\,$cSt
\\
Density (silicone oil)       
& $\rho$        
& --
& $0.96 \,$g/cm$^3$
\\
Surface tension (silicone oil)  
& $\sigma$        
& --                            
& $20.5 \,$dynes/cm
\\
Kinematic viscosity (silicone oil)  
& $\nu$ 
& --
& $5 \,$cSt
\\
Gravitational acceleration
& $g$
& --
& $981\,$cm/s$^2$
\\
Weber Number
& $\We$     
& $\rho V_0^2 R/\sigma$ 
& 
$0.5-8.0$
\\
Bond Number 
& $\Bo$
& $\rho g R^2/\sigma$
& 
$0.017-0.056$
\\   
Ohnesorge Number
& $\Oh$ 
& $\mu/\sqrt{\sigma R \rho}$  
& 
$0.006-0.057$
\\
Reynolds Number
& $Re = \sqrt{We}/Oh$ 
& $ \rho V_0 R /\mu$  
& 
$15-280$
\\

\end{tabular}
\label{tab:parameters}
\end{table}

\subsection{Experimental procedure}\label{Section:ExpProc}
Care must be taken to ensure that both the fluid interface and the fluid in the reservoir of the drop generator are contaminant-free, as dust or surfactants can modify the physics involved. Prior to each experiment, the bath and tubing are thoroughly cleaned with an isopropyl alcohol solution, flushed with deionized water, and then left to dry in a fume hood with particulate filtering for 30 minutes. The drop generator and nozzle are cleansed with an ethanol solution, and then flushed with deionized water for 5 minutes. Gloves are worn at all times to minimize contamination. The drop generator is controlled by an Arduino Uno board, with a simple H-bridge circuit initiating the voltage switching of a DC power supply \citep{ionkin2018note}. The fluid within the bath is periodically flushed, approximately after every 15 droplet impacts to reduce surface contamination \citep{kou2008method}. Overflow from the flushing is caught by a small lip in the bottom of the bath, which is then drained to a waste container. There are two syringes connected to the bath, which allow for fine adjustments of the equilibrium bath depth after flushing.

We collect experimental data for the top and bottom of the droplet during free flight. During contact, we track the height of the top of the droplet and that of the center of the deformed free surface. Since the air layer that separates the droplet and the interface is negligibly thin relative to the scale of the droplet, we assume that this point is also effectively the location of the droplet's south pole. Just after the drop rebounds off the surface, the axisymmetric surface wave created by the impact is partially in the line of sight of the camera, and obscures the bottom of the droplet for a brief period during take-off. These data points have been omitted from the bottom trajectory when reported. The raw video data are post-processed using a custom Canny edge detection software implemented in MATLAB 2021b, which quantifies the droplet trajectory, and then computes impact parameters and bouncing metrics \citep{galeano2021capillary}. 

There are several metrics of interest in our study, which we define in what follows. The maximum penetration depth, $\delta$, of a bounce is defined as the position of the bottom of the droplet at the lowest point in the trajectory (relative to the undisturbed interface height).  In our experiment, the contact time, $t_c$, is defined as the time duration from which the top of the droplet crosses the height $z=2R$ to the time the top of the droplet returns to that height.  Due to the nature of visualisation setup it was impossible to determine precisely when the droplets lost physical contact with the fluid; however, this always occurred before the top of the drop returned to the level $z=2R$.  Each bounce was also characterised by its coefficient of restitution, $\alpha$, which is defined here as the negative of the normal exit velocity, $V_e$, divided by the normal impact velocity, $V_0$. This parameter ranges between 0 and 1, and is related to the momentum exchange during impact.  

In order to determine the contact time ($t_c$) and coefficient of restitution ($\alpha$), a parabola was fitted using a least-squares method to the incoming and outgoing trajectories, separately, with at least 30 data points prior to impact and at least 40 data points following rebound.  The analytical form of the parabolic fit was then used to extrapolate the time at which the sphere crosses the still air-water interface height (which corresponds to a root of the parabolic function).  The derivative of the parabolic fit function was then computed analytically and its value evaluated at these times in order to calculate the impact speed, $V_0$, and exit speed, $V_e$ \citep{galeano2021capillary}.   
\section{Linearized quasi-potential fluid model}
\hspace*{1em} In this section, we develop a model for droplet impact on a flat fluid interface from first principles. First, we use the linearized Navier-Stokes equations to model the flow within the bath and the shape of the bath interface. Then, we use an orthogonal function decomposition of the bath model to derive a single set of linear ordinary differential equations (ODEs) that govern the bath mode amplitudes.  We then write a similar model for the droplet interface, which reduces to another set of linear ODEs governing droplet mode amplitudes.  Finally we propose a model for the pressure distribution and its extent acting on the bath and droplet during contact, couple the two sets of ODEs together using a single-point kinematic match condition, and solve the system implicitly using standard numerical integration techniques.  A schematic of the problem is illustrated in Figure \ref{fig1:setup}(d).

\subsection{Bath interface model}
The present work models the bath interface dynamics using a linearized, quasi-potential flow model following the work of \cite{galeano2017non}. For the problem of a droplet impacting on a free interface, the Navier-Stokes equations govern the flow generated by the bath-droplet interaction. Assuming the flow to be incompressible, isothermal, and Newtonian, we can define the fluid velocity vector $\textbf{u} = [u,v,w]^T= \nabla \phi + \nabla \times \Psi$ and the bath interface shape $\eta =\eta(r,\Theta,t)$,  Here, $\phi$ is the scalar potential and $\Psi$ is the vector stream function. We then linearize the governing equations and boundary conditions about the undisturbed free surface $z=0$. Utilizing the arguments presented in \cite{galeano2017non} and \cite{dias2008theory}, we can recast the governing equations to be
\begin{align}
    \nabla^2 \phi + \partial_z^2 \phi &= 0 \text{,\hspace*{1em}} z \leq 0 \text{,} \label{potential}\\
    \partial_t \eta  &= \partial_z \phi + 2 \nu \nabla^2 \eta \text{,\hspace*{1em}} z = 0 \text{,} \label{kbcsum}\\
    \partial_t \phi &= - g\eta -2\nu \partial_z^2 \phi + \frac{\sigma}{\rho}\kappa - \frac{p_s}{\rho} \text{,\hspace*{1em}} z = 0 \text{,} \label{dbcsum}\\
    \partial_z \phi &=  0 \text{ at } z=h_0 \text{.} \label{final_equations}
\end{align} Here, $p_s(r,\Theta,t)$ is the contact pressure, $g$ is the gravitational acceleration, $\kappa(r,\Theta,t)=\nabla^2 \eta$ is twice the linearized mean curvature of the interface, $h_0$ is the depth of the undisturbed bath, and $\rho$, $\sigma$, $\nu = \mu/\rho$ are the fluid density, surface tension, and kinematic viscosity, respectively. 
In this notation, $\partial_{( )}$ denotes partial differentiation with respect to the variable given in the parenthesis and $\nabla^2 = \partial_{r}^2 + (1/r)\partial_r + (1/r^2)\partial_{\Theta}^2$. The tangential stress boundary conditions are automatically satisfied in these approximations.  As detailed in \cite{galeano2017non}, this leading order theory is valid in the weakly viscous limit when $Re=\sqrt{We}/Oh \gg 1$.  A similar bath model was used by \cite{blanchette2016modeling,blanchette2017octahedra}, although the viscous correction term was not included in the dynamic boundary condition (\ref{dbcsum}).\\
\hspace*{1em} We assume that the impact occurs in a bath of some viscous fluid which is subject to two boundary conditions, $\partial_n \phi = 0$ on the walls of the bath and $\partial_z \phi =0 $ on the bottom of the bath, where $n$ is the outward facing normal of the walls of the bath \citep{benjamin1954stability}. The former condition implies that $\partial_n \eta =0$ on the walls of the container.  These conditions correspond physically to a bath where the working fluid maintains a constant contact angle of $90^{\circ}$ at the walls, and has no-flux boundary conditions along the walls and bottom. Applying these conditions to the governing system of equations, an orthogonal function expansion for the unknowns $\eta$, $\phi$ and their derivatives can be explicitly written. \\ \hspace*{1em} The orthogonal basis functions $S_{j,m}(r,\Theta)$ ultimately  must satisfy \begin{equation}
    (\nabla^2 + k_{j,m}^2) S_{j,m} = 0 \text{, }
\end{equation}
inside of any bounding curve $ K $ that exists on the border of the bath and the boundary condition of $\partial_n S_{j,m} = 0$ on $K$.  $k_{j,m}$ are the eigenvalues of the system, and depend on the choice of the physical domain of the problem. If the boundary curve $K$ is a circle, the functions become $S_{j,m} = J_j(k_{j,m} r) \cos{j \Theta}$.  Here $J_j$ are Bessel functions of the first kind and $k_{j,m}$ are the solutions to $J'_j(k_{j,m}b) = 0$, where $b$ is the bath radius. If we further assume the problem to be axisymmetric, then we can choose $j=0$, and write $S_{0,m} = J_0(k_{0,m} r)\text{.}$  For convenience, we define $k_{0,m}=k_m$ henceforth.  We then can express the free surface elevation as 
\begin{equation}
    \eta(r,t) = \sum_{m=0}^{\infty} a_m(t) J_0(k_{m} r) \text{.}
\end{equation}
We then re-write all of the unknowns of the axisymmetric bath problem, using (\ref{potential}) and (\ref{kbcsum}), as a function of the time varying amplitude coefficients, $a_m(t)$:
\begin{align} 
    \eta(r,t) = &\sum_{m=0}^{\infty} a_m(t) J_0(k_m r) \text{,} \label{bathmodelsum1}\\
    \kappa(r,t)=\nabla^2 \eta = -&\sum_{m=0}^{\infty} k_m^2 a_m J_0(k_m r) \text{,} \label{bathmodelsum2}\\ 
    \phi(r,z,t) = &\sum_{m=0}^{\infty} \left( \frac{d a_m}{dt} + 2 \nu k_m^2 a_m \right) J_0(k_m r) \frac{\cosh{k_m(h_0+z)}}{k_m\sinh{k_m h_0}} \text{.} \label{bathmodelsum3}
\end{align}
\hspace*{1em} In order to arrive at the final equations of motion for the free surface, we take the decompositions (\ref{bathmodelsum1}-\ref{bathmodelsum3}) and substitute them into the dynamic boundary condition (\ref{dbcsum}).
Rearranging, we find 
\begin{equation}
    \sum_{m=0}^{\infty} \left [ \frac{d^2 a_m}{dt^2} + 4 \nu k_m^2 \frac{da_m}{dt} + \left(\frac{\sigma k_m^2}{\rho} + g\right)k_m\tanh{(k_m h_0)}a_m \right ] J_0(k_mr) = -\frac{p_s}{\rho}k_m\tanh{(k_m h_0)} \text{.}
    \label{bathmodeswithP}
\end{equation}
Each wave mode in the bath is described by a forced, damped harmonic oscillator equation.
\subsection{Droplet interface model}
Additionally, we wish to recover a similar set of equations that describe the gravity-capillary waves present in the droplet, and then couple these equations to the motion of the bath.  The full derivation of the droplet oscillation model can be found throughout prior works \citep{lamb1924hydrodynamics,tsamopoulos1983nonlinear,courty2006oscillating,chevy2012liquid,balla2019shape} and is briefly summarized below.

We begin by utilizing spherical harmonics to decompose the droplet radius in a spherical domain, 
\begin{equation}
    \xi(\theta,t) = R + \sum_{l=1}^{\infty} \sum_{n=-l}^l \beta_l^n(t) Y_l^n(\cos{\theta},\phi) \text{,}
    \label{xidefined}
\end{equation}
with $Y_l^n =  P_l^n(\cos{\theta}) e^{i n \phi}$.
Due to the axisymmetry of the problem, we set $n=0$ and the spherical harmonics reduce to associated Legendre polynomials, $P_l^n (\cos(\theta))$. For convenience, we write $\beta_l^n=\beta_l$ henceforth. We assume that the velocity potential takes the same form as the decomposition of the interface \citep{lamb1924hydrodynamics, balla2019shape}.
We then turn to an energy conservation equation of the form
\begin{equation}
    \frac{d T}{dt} = \dot{W} - \epsilon \text{,}
\end{equation}
with $T = K + G$, $\dot{W}$, and $\epsilon$, as the total energy (sum of the kinetic energy $K$ and potential energy $G$) of the drop, the rate of work done on the droplet interface, and the viscous dissipation, respectively. We can express $T$, $\dot{W}$, and $\epsilon$ using the decomposition (\ref{xidefined}), and substitute these expressions into the conservation of energy equation. Then, utilizing the linearized kinematic boundary condition yields a set of forced, damped harmonic oscillators that describe the amplitude of each individual spherical mode,
\begin{equation}
     \sum_{l=1}^{\infty} \left[ \frac{d^2 \beta_l}{dt^2} + 2 \alpha_{l} \frac{d \beta_l}{dt} + \omega_{l}^2 \beta_l \right ] = \sum_{l=1}^{\infty} \left[ -\frac{(2l+1)l}{2\rho R} \int_0^{\pi} p_s(\theta) \sin{\theta} P_l(\cos{\theta}) d\theta + g \delta_{1l} \right ] \text{.}
\end{equation}
We drop the sums, and arrive at the result
\begin{equation}
     \frac{d^2 \beta_l}{dt^2} + 2 \alpha_{l} \frac{d \beta_l}{dt} + \omega_{l}^2 \beta_l = -\frac{(2l+1)l}{2\rho R} \int_0^{\pi} p_s(\theta) \sin{\theta} P_l(\cos{\theta}) d\theta + g \delta_{1l} \text{,}
    \label{dropmodeeq}
\end{equation}
with \begin{align}
    \alpha_{l} &= (2l+1)(l-1)\frac{\mu}{\rho R^2} \text{,}\\
    \omega_{l}^2 &= l(l-1)(l+2) \frac{\sigma}{\rho R^3} \text{,}
\end{align}
and $\delta_{1l}$ is the Kronecker delta function.  This model is valid in the weakly viscous limit, when $Oh\ll1$.
An extension of the free droplet model to arbitrary $Oh$ can be found in other prior works \citep{chandrasekhar2013hydrodynamic,molavcek2012quasi,miller1968oscillations}.
\subsection{Pressure forcing during impact}
\label{subsec:modelPressure}
There is still an additional unknown in the bath mode (\ref{bathmodeswithP}) and drop mode (\ref{dropmodeeq}) equations: the applied pressure distribution $p_s(r,t)$. This is generally a function of the properties of the fluid medium, the impacting speed of the object, the shape of the impacting object, and the motion of the gas that surrounds the fluid. However, in this work, we will assume that the viscosity of the ambient gas is small relative to the fluid bath such that the flow within the small air film is negligible, and the pressure acts solely to apply an upwards hydrodynamic force on the droplet. In non-dimensional terms, we can construct two additional restrictions for our model, $\rho_g / \rho \ll 1$ and $\Oh_g = \mu_g / \sqrt{\rho \sigma R} \ll \Oh$ following prior work \citep{molavcek2012quasi}. In the current experimental and direct numerical simulation work, $\rho_g/ \rho \sim \mathcal{O}(10^{-3}) $ and $\Oh_g  \sim \mathcal{O}(10^{-4})$, thus the influence of these two additional parameters is indeed negligible.  We have verified this for our typical experimental parameters through DNS, and find that both a 4-fold increase and decrease in the ambient density and viscosity from air properties at standard temperature and pressure (STP) produces negligible changes to the trajectory, instantaneous shape of the droplet, and free surface shape throughout the interaction of the droplet and bath (Appendix A). 

The radial extent of the pressure distribution is generally unknown for impact problems, and constitutes an additional problem that we must solve. In the present work, we assume that this unknown pressure distribution takes the form 
\begin{equation}
    p_s(r,t) = \frac{F(t)}{\pi R^2} H_r(r/r_c(t)) \label{Presdef}
\end{equation}
where $F$ is the instantaneous magnitude of the contact force, evaluated at $r=0$ and $H_r$ is an assumed spatial profile of the pressure in the contact region.  For this distribution, we can use a function that resembles the true shape of the pressure distribution during contact. The contact region, $A_c$, will be assumed to a simply-connected disk, following \cite{galeano2017non} and \cite{korobkin1995impact}. This allows us to write a single unknown $r_c(t)$ to fully describe the temporal evolution of this region of contact. \cite{blanchette2016modeling,blanchette2017octahedra} used a fixed parabolic pressure shape function

\begin{equation}
H_r(r) =  \begin{cases} 
      C \left( 1 - (\frac{r}{R})^2 \right) & r\leq R \\
      0 & r > R \text{.}
   \end{cases}
\end{equation}

Here, $C$ is the constant magnitude of the pressure at $r=0$ and $R$ is the undeformed radius of the droplet.  \cite{blanchette2016modeling,blanchette2017octahedra} chose the value of the magnitude $C$ such that $\int_0^b p_s r dr= \pi R^2$. Thus, the pressure acting on the bath interface in the respective models had a constant pressure shape function $H_r$ for all times during contact. However, simulation results from \cite{galeano2017non,galeano2021capillary} show that the shape of the pressure distribution at the surface of a fluid bath due to an impacting, non-wetting sphere is flatter and more similar to a top-hat function for most times, and that the spatial extent of the distribution changes continuously with time during impact. Additionally, for a droplet impacting on a solid surface, the pressure in the air film has been inferred by \cite{de2015wettability}. The air film thickness during a bounce was measured using interferometry and the pressure estimated using a lubrication model. The film pressure in both the impacting and rebounding regimes is approximately uniform, with deviations from uniformity only near the edge of the film. In related work, the impact pressure between a droplet and a wettable solid substrate has been studied extensively by \cite{mandre2009precursors} and \cite{mani2010events}, and their results indicate that the impact pressure increases sharply near the contact line, likely a consequence of the decreased air film thickness in that region.  For the case of droplets bouncing on a deep pool, \cite{tang2019bouncing} measured the air film thickness and found the film thickness to be significantly more uniform in both impacting and rebounding stages for $\We$ values similar to those explored in the present work, presumably as a result of the deformability of the substrate and impactor.  Our predictions from direct numerical simulation (presented and discussed in section \ref{sec.ModellingAndSimulations}) similarly suggest a more uniform air film thickness for the present problem, and a nearly uniform pressure profile during all stages of rebound. \\
\hspace*{1em} We will use a simple polynomial that resembles a smoothed top hat in this work, with 

\begin{equation}
H_r(r/r_c(t)) =  \begin{cases} 
      C \left( 1 - (\frac{r}{r_c(t)})^6 \right) & r\leq r_c(t) \\
      0 & r > r_c(t) \text{.}
   \end{cases}
\end{equation}

In order to remain consistent with our linearization, we do not allow $r_c$ to exceed $R$.  Requiring that the integral of the pressure over the contact area is $F(t)$, we find
\begin{equation}
   \int_0^{b} H_r(r/r_c) r \ dr = \frac{R^2}{2} \text{,}
\end{equation}
which sets the constant $C$ in the pressure shape function $H_r(r/r_c)$. Our bath model relies on the decomposition of the fluid motion into a linear superposition of infinitely many waves with wavenumbers $k_m$. Therefore, we apply a similar decomposition to this pressure function $p_s=\sum_{m=0}^{\infty} d_m J_0(k_m r)$. Since we are working in a cylindrical domain, we will choose the zeroth order Bessel function of the first kind as our orthogonal basis function, and thus the $d_m$ are the Fourier-Bessel coefficients of the function $H_r$,
\begin{equation}
    d_m = \frac{2}{(b J_1(k_m))^2}\int_0^b H_r(r) r J_0(k_m r) dr, 
\end{equation} with the domain extending from $r=[0,b]$. The reconstruction of the top-hat function in Fourier-Bessel space converges too slowly to be of practical use \citep{storey1968convergence}, also noted by \cite{blanchette2016modeling}, and as such we use a polynomial expression that resembles a smoothed top hat. Additionally, we tested higher order polynomials (corresponding to a larger flat region), and found increasingly poor convergence behavior, similar to that of the top hat (see Appendix B for a case study on the sensitivity of the results to the choice of shape function).  The ultimate choice of shape function used here thus represents a practical compromise.\\
\hspace*{1em} Substituting in the definition of the pressure (\ref{Presdef}) into (\ref{bathmodeswithP}), performing the Fourier-Bessel decomposition, we find
\begin{equation}
    \frac{d^2 a_m}{dt^2} + 4 \nu k_m^2 \frac{da_m}{dt} + \left(\frac{\sigma k_m^2}{\rho} + g\right)k_m\tanh{(k_m h_0)}a_m = -\frac{F}{\rho \pi R^2} d_{m} k_m\tanh{(k_m h_0)} \text{,}
\end{equation}
which govern the evolution of bath wave modes $m$.
Similarly, substituting the pressure (\ref{Presdef}) into (\ref{dropmodeeq}), we write
\begin{equation}
    \frac{d^2 \beta_l}{dt^2} + 2 \alpha_{l,0} \frac{d \beta_l}{dt} + \omega_{l,0}^2 \beta_l = -\frac{F}{2 \pi \rho R^3} c_{l} (2l+1) l + g \delta_{1l} \text{.}
    \label{dropmodeeq2}
\end{equation}
The coefficients $c_{l}$ result from the mode decomposition of the projection of the pressure into spherical space,
\begin{equation}
    c_{l} = \int_0^{\pi} H_r(\theta) \sin(\theta) P_l(\cos{\theta}) d\theta  \text{,}
\end{equation}
which naturally arise in the derivation of equation (\ref{dropmodeeq}). 
Additionally, the definition of the pressure (\ref{Presdef}) reduces the governing equation of the $l=1$ ``translational'' mode to
\begin{equation}
    \frac{d^2 \beta_1}{dt^2} = -\frac{F}{m} + g \text{, } 
\end{equation}
which clearly governs the droplet center of mass motion $\beta_1=-z_{cm}$.
Evidence from the simulations of \cite{galeano2017non} indicates that impact trajectory is very sensitive to the instantaneous size of the contact area. Utilizing a constant pressed area for the pressure, as done in \cite{blanchette2016modeling}, does not produce results that compare well with experiment (Appendix B), particularly for cases of small $We$. Figure \ref{fig:pressuredrop} in Appendix B depicts how the choice of this pressure shape function modifies the predicted trajectory of the droplet for typical experimental parameters. The trajectory is largely insensitive to the choice of pressure shape function, but incorporating a time-dependent contact radius is essential for agreement.  The method for determining both $F(t)$ and $r_c(t)$ are discussed in the next section.

\subsection{Modeling contact}
The contact force $F(t)$ is determined through the use of a ``1-Point'' kinematic match (1PKM) condition. Essentially, we enforce contact only at a single point; the center of our axisymmetric domain. Thus, the additional constraint can be written as
\begin{equation}
    \eta(r=0,t) = z_{cm}(t) - \xi(\theta=0,t) = \sum_{m=0}^{\infty} a_m(t) = z_{cm}(t) -\left(R + \sum_{l=2}^{\infty}\beta_l(t)\right) \text{.}
\end{equation}
This additional constraint allows us to determine the unknown contact force $F(t)$. Contact between the droplet and the bath ends when the magnitude of the contact force as predicted by the kinematic match becomes negative. We note that this 1PKM model is a significant simplification of the full kinematic match successfully used to study related impact problems \citep{galeano2017non,galeano2019quasi,galeano2021capillary}. 
The full kinematic match predicts the evolution of the
contact area and the contact pressure distribution (without requiring an assumption for $H_r$) by imposing natural geometric and kinematic constraints, essentially considering additional equations to solve at each time step. The algorithm requires iteration at each time step, and the minimization of a tangency boundary condition is used to determine the correct contact area and pressure shape.

Lastly we turn to the unknown contact radius $r_c(t)$.  By not restricting the deformations of the bath and droplet interface with the use of additional tangency and distributed kinematic match conditions, we find that the results of our simulation consistently produce interfacial shapes that cross each other. The amount of overlap between the two interfaces is generally small, for instance, in the comparison in Figure \ref{fig:traj} the maximum overlap is less than $0.05R$.    However, we can use the predictions from both interfacial models to determine the exact location where the two interfaces cross and separate, and use this as the instantaneous radius of contact $r_c(t)$. Thus, at each time, contact between the bath and drop is ensured at both $r=0$ and $r=r_c(t)$.  In order to enforce contact within the entirety of the contact region, a full kinematic match would be required - this circumvents the need for any assumptions on the pressure profile shape, but is substantially more computationally expensive. Our contact radius criterion is similar to that of the numerical model presented in prior work on droplet rebound from solid substrates \citep{molavcek2012quasi}. While this method is unphysical, it yields accurate predictions for the contact radius as compared to DNS (Figure \ref{fig:traj}(b)).

\subsection{Summary}
Choosing a timescale of $t_{\sigma}=\sqrt{\rho R^3/\sigma}$ (with $\tau = t/t_{\sigma}$), a length scale of $R$, and a force scale of $2 \pi \sigma R$ (with $f = F / 2 \pi \sigma R )$, we  recast the governing equations in non-dimensional form as 
\begin{align}
    \eta(r,\tau) = \sum_{m=0}^{M} a_m(\tau) J_0(k_m r) \text{,} \label{bathshape} \\
    \xi(\theta,\tau) = 1 + \sum_{l=2}^{L}\beta_l(\tau) P_l(\cos{\theta}) \text{,} \label{dropshape} \\
    \frac{d^2 {a_m}}{d\tau^2} + 4 {\Oh} {k_m}^2 \frac{d {a_m}}{d\tau} + \left( {k_m}^2 + \Bo \right){k_m}\tanh{({k_m} {h_0})}{a_m} =& -2{f} d_{m,0} {k_m} \tanh{({k_m} {h_0})} \text{,} \label{bathinterfaceFin} \\
    \frac{d^2 {\beta_l}}{d\tau^2} + 2 \Oh (2l+1)(l-1) \frac{d {\beta_l}}{d\tau} + l(l-1)(l+2) {\beta_l} =& - {f} c_{l} (2l+1)l + \Bo\delta_{1l} \text{,} \label{dropinterfaceFin} \\
        \frac{d^2 z_{cm}}{d \tau^2} = \frac{3}{2} f - \Bo \text{, } \label{centerofmass}\\
    \eta(0,\tau) =  {z}_{cm}(\tau) -  {\xi}(\theta=0,\tau)  \text{.} \label{1pkm} 
\end{align}
Equations (\ref{bathinterfaceFin}) and (\ref{dropinterfaceFin}) describe the evolution of the bath and droplet oscillation modes, respectively. Equation (\ref{centerofmass}) governs the vertical motion of the droplet's center of mass.  Equation (\ref{1pkm}) couples these equations all together, with $r=0$ as the single point of ``contact'' enforced between the droplet and the bath and allows for determination of the unknown $f(\tau)$. These equations are solved using standard ordinary differential equation numerical integration techniques.  The shape of the bath and droplet can be reconstructed at any time $t$ via the sums in equations (\ref{bathshape}) and (\ref{dropshape}), respectively.

The complete model is valid when $Re = \sqrt{We}/Oh \gg 1$ and $Oh\ll1$.  Also, since the model is linearized about the undeformed state, we anticipate it to hold when deformations remain small, further suggesting $Bo\ll1$ and $We\ll1$.  However, we later demonstrate through direct comparison with experiment and DNS that the model remains predictive even for moderate $We$. 
\subsection{Numerical methods}
\hspace*{1em} We solve these equations using a backward Euler method, ensuring a minimum of 100 time steps within the inertio-capillary time $t_{\sigma}$. An implicit method was chosen, following \cite{galeano2021capillary}, as the instantaneous size of the pressure distribution acting on both the droplet and bath at the next time step is unknown. Treating the pressure explicitly on either the droplet or the bath can lead to nonphysical behavior in the system. We used $M=150$ modes for the bath interface and $L=55$ modes for the droplet interface. These values were determined by running simulations of a $\We$ $=0.7$, $\Bo$ $=0.017$, $\Oh$ $=0.006$ impact and assessing convergence as described in what follows. First, we kept the number of droplet modes fixed at $L=15$ and increased the number of bath modes from $30$ to $500$ in increments of $25$. Then, the simulation was run again, fixing the number of bath modes at $75$ and increasing droplet modes from $15$ to $200$. Sufficient convergence was determined if the maximum absolute value of the difference in center of mass trajectories during contact changed by less than $1\%$. Finally, both the droplet and bath number of modes were increased simultaneously, and convergence was still observed. These values are similar to comparable to those found in \cite{blanchette2016modeling} ($M=200$, using sine functions as the basis functions in a square bath), and \cite{molavcek2012quasi} ($L=150$, but found good accuracy in comparison to experimental data at $L=20$). Additionally, once mode convergence was determined, we decreased the time step of the simulation in increments until time step convergence was similarly reached using the same criterion. Unexpectedly, in a fully-converged simulation, there is a time step threshold below which the algorithm results in unstable oscillations in the magnitude of $F(t)$. This time step threshold is typically at least three orders of magnitude smaller than $t_{\sigma}$. This apparent instability deserves awareness and future attention, but does not affect the results presented in the present work.  The bath size was set to $b=25 R$ which was determined to be sufficiently large such that reflected waves did not influence the droplet during impact. In order to find the instantaneous contact radius, we take two line segments from the reconstruction of the droplet and bath interface shapes. From these we can write a linear system of equations for four unknowns: the $(r,z)$-pair of the intersection location, the normalized distance from the starting point of the first line segment to the intersection, and the normalized distance from the starting point of the second line segment to the intersection \citep{intersections}. 
We then loop over every line segment to find every intersection. We take the largest of this set as $r_c(t)$. In the reconstructions of the interfaces at each time step, at least 5000 points are used in both $\theta$ and $r$ to ensure that error is minimized.  All code associated with the implementation of the model is  available at  \url{https://github.com/harrislab-brown/BouncingDroplets}.
\section{Direct numerical simulation}\label{sec.ModellingAndSimulations}
 
Complementing the previous investigative tools in terms of both experimental and modelling capabilities, we build a dedicated computational framework within the open-source solver Basilisk. This has the dual aim of validation and exploration of quantities of interest outside the reach of previous methodologies. Basilisk \citep{popinet2015quadtree} and its predecessor Gerris \citep{popinet2003gerris, popinet2009accurate} have been widely adopted by the computational fluid dynamics community over the past two decades. Its second-order accuracy in both space and time, alongside adaptive mesh refinement and parallellisation features, have led to successful studies of multi-scale fluid systems such as the scenario here. The equations for conservation of momentum and mass are solved as part of a one-fluid formulation within the volume-of-fluid (VOF) methodology, with the Bell-Collela-Glaz advection \citep{bell1989second} employed for advective terms using a CFL-limited timestepping strategy, a well-balanced surface tension implementation \citep{popinet2018numerical} and an implicit treatment of the viscous terms.

\begin{figure}
    \centering
    \includegraphics[width=1\textwidth]{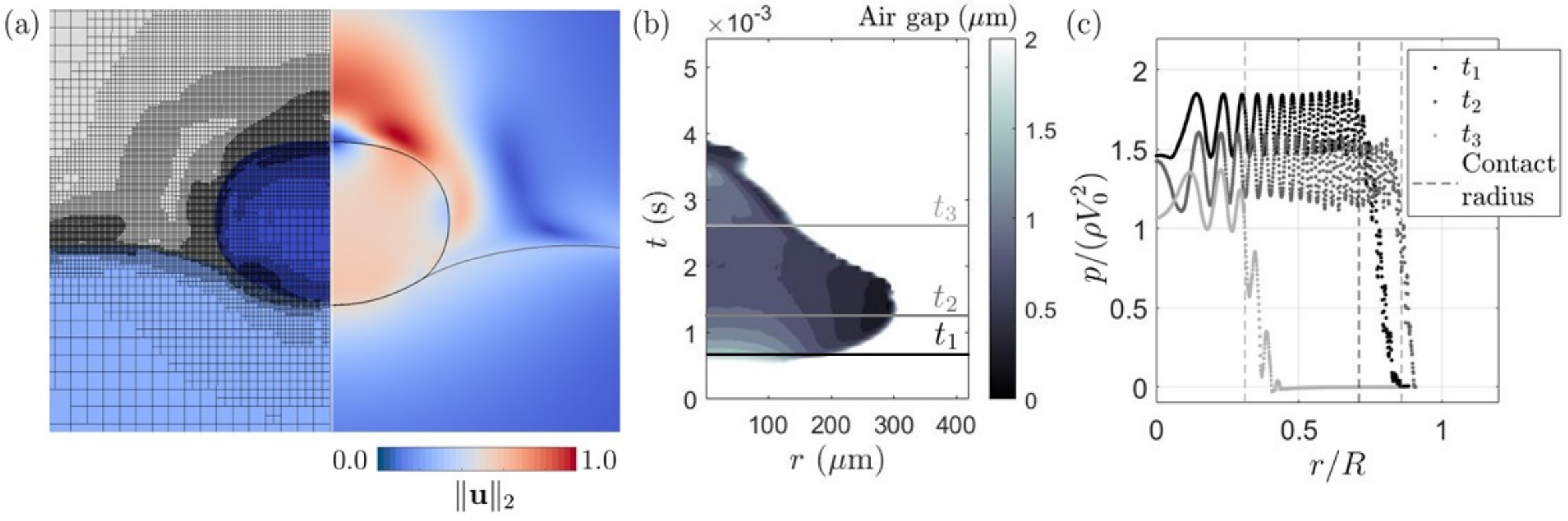}
    \caption{Direct numerical simulation detail of the rebound of a $R=0.35$ mm deionized water droplet from a bath of the same fluid corresponding to $\We$$=0.7$, $\Bo$$=0.017$, and $\Oh$ $= 0.006$. (a) Adaptive mesh refinement strategy overlaid to highlighted drop, pool and gas regions (left); $2$-norm of the velocity field $\|\textbf{u}\|_2$ and fluid-fluid interfaces (right) at dimensionless $t=1.5$. (b) Thickness of the air gap entrapped between the impacting drop and the pool, with three highlighted timesteps during descent, maximal spread, and rebound, presented in dimensional form. (c) Raw pressure data extracted at the centre of the air gap, along the contact radius, for the highlighted timesteps in panel (b). Video supplementary material containing additional detail is also provided.}
    \label{fig:DNSexample}
\end{figure}

The non-dimensionalisation described in Table~\ref{tab:parameters} is retained, with the drop diameter $R$ and initial impact velocity $V_0$ providing the reference lengthscale and velocity scale in the system. The additional gas region, fully captured in the DNS, has physical properties modelled from typical air values at room temperature. In particular we highlight the value of density ratio $\rho_g/\rho \approx 0.0012$ and $\Oh_g = O(10^{-4}) \ll \Oh$, as outlined in Table~\ref{tab:parameters}, both consistent with the modelling assumptions described in subsection~\ref{subsec:modelPressure}. The axisymmetric computational domain is constructed as a $20R \times 20R$ square, with $20R$ proving sufficient to avoid any artifacts from boundary reflections, while describing the dynamics of the impact \citep{galeano2021capillary}. The pool height is set to $z=10R$, rendering bottom effects negligible, while also allowing a generous region occupied by air for the droplet bounce to be quantified. Given that scales range from microns (in the gas layer between drop and pool) to centimetres (for the full domain size), this is a challenging setup which takes full advantage of the numerical capabilities available. The adaptive mesh refinement strategy prioritises fluid-fluid interfacial location and changes in the magnitude of the velocity components in order to concentrate resources where needed. Extensive validation efforts have determined that a minimum grid cell size of approximately $3\ \mu$m (or just above $100$ cells per radius) is sufficient to ensure mesh independence for the observed metrics, with additional refinement having been further tested in sensitive regimes and yielding no substantial benefit. This leads to a grid cell count of $20000-80000$ over the duration of a simulation, with the workload typically distributed across 4-8 cores over approximately $48$ CPU core hours per run. A subset of the computational domain, including the grid cell distribution, is shown in Figure~\ref{fig:DNSexample}(a). We note that, with a uniform mesh, a similarly resolved computation would require in excess of $4.2$ million grid cells, rendering full parameter studies intractable. Similar deployment of resources has previously proven successful in the investigation of multi-fluid systems in the context of impact regimes ranging from small \citep{galeano2021capillary} to moderate \citep{fudge2021dipping} and finally large \citep{cimpeanu2018early} velocities, resulting in tools and improved insight into physical phenomena such as bouncing, coalescence and splashing. The implementation described above is made available to interested users at \url{https://github.com/rcsc-group/BouncingDroplets}.

A particularity of our setup lies in the integration of the recent functionality developed for non-coalescence scenarios, as previously employed by \cite{ramirez2020lifting} and recently by \cite{sanjay2022taylor}. With different symbolic definitions for underlying colour functions representing the droplet and the bath in the VOF framework, the algorithm ensures that numerically induced coalescence is avoided, and the entrapped gas region between the impacting drop and the pool is well resolved throughout the studied motion. A sufficiently high resolution level is still required for $\mathcal{O}(1)\ \mu\textrm{m}$ thicknesses to be maintained \citep{tang2019bouncing}, as illustrated in Figure~\ref{fig:DNSexample}(b), with the non-coalescence package allowing suitable sub-grid cell level lengthscales to be reached. The region highlighted in colour represents the approximate contact area as a function of space and time, with the contact radius limit defined by a gap of twice the typical width of the gas film trapped between the drop and pool being reached as one navigates radially outwards (a definition consistent with the mathematical model). The typical gas film profile in space is given by a nearly uniform value in the bulk of the drop-pool contact area, with a thinning observed towards the very end, and the smallest scales reached therein approaching $0.25\ \mu$m. While a direct comparison with existing experimental data would require a more dedicated setup and is beyond the scope of this study, similar lengthscales were recently observed by \citet{tang2019bouncing}. Finally, this study also constituted one of our most stringent convergence tests, with more refined mesh levels producing only negligible changes in this observed space-time map. The time variation of the contact radius is a clear hallmark of this behaviour, with an initially rapid increase to almost $85\%$ of the initial droplet radius being followed by a slower decrease as restorative forces push the droplet back up during the ascent stage, consolidating this as a key assumption to be retained as part of the mathematical model presented previously. While the above discussion focuses on a representative set of parameters, the generated outcomes and insight apply to the much wider parameter set we explore.

Figure~\ref{fig:DNSexample} (accompanied by video supplementary material) also outlines velocity field information inside each of the fluid phases (a-right), as well as the aforementioned gas film thickness (b) and pressure distribution (c) across relevant flow stages. Manipulating raw pressure data in projection methods is a known challenge \citep{philippi2016drop,negus2021droplet}, with the oscillatory behaviour observed in Figure~\ref{fig:DNSexample}(c) linked to the VOF approximation on a structured grid, and the underlying colour function sampled across various approximation points inside the cells in near vicinity of interfaces with large value contrasts in the quantities of interest. We underline in particular that the aforementioned variations oscillate around well-defined means within the contact region. The pressure in the air gap across the evolving contact radius may thus be robustly approximated by a top-hat function at every flow stage investigated, reinforcing previous modelling choices in subsection~\ref{subsec:modelPressure} and Appendix~\ref{app:pressureShape}.

\begin{figure}
    \centering
    \includegraphics[width=1\textwidth]{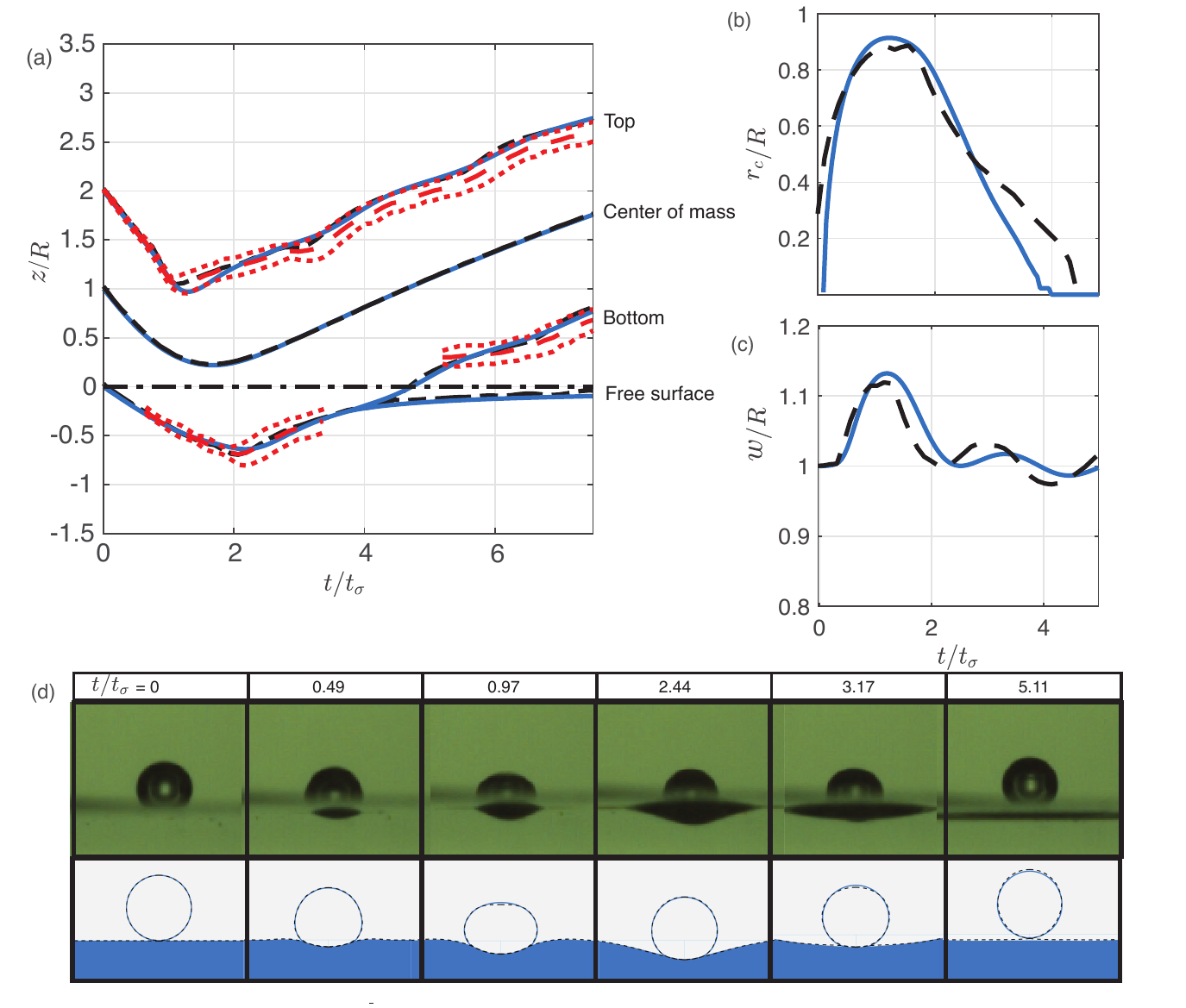}
    \caption{Rebound of a $R=0.35$ mm deionized water droplet from a bath of the same fluid corresponding to $\We$$=0.7$, $\Bo$$=0.017$, and $\Oh$ $= 0.006$. (a) Trajectory comparison between the experiment (red dashed line with typical variation shown as dotted red lines), DNS (black dashed line), and quasi-potential model (blue solid line). (b) Instantaneous contact radius, normalized by the undeformed radius $R$, as a function of time for the quasi-potential model and DNS.  (c) Maximum width of the droplet $w$, as a function of time for the quasi-potential model and DNS. (d) Comparison of droplet shape between experiment, DNS, and quasi-potential model.  Video supplementary material is available.}
    \label{fig:traj}
\end{figure}

After having undergone stringent numerical verification procedures, a typical study in our parameter regime of interest is showcased in Figure~\ref{fig:traj} and described in detail in the following section. The agreement between experiment, model and simulation is very encouraging, with the developed resources in an excellent position to bridge one another and comprehensively explain the target rebound metrics.

\section{Results} 
In this section, we first present the results of a direct comparison between experiment, quasi-potential model, and DNS for a single impact $\We$. Then, we vary $\We$ for two working fluids, and compare the results of the three different impact metrics between the experiment, DNS, and model.  Having validated the model and DNS, we then run sweeps over $\Bo$ and $\Oh$ to deduce the effect that these non-dimensional constants have on droplet rebound metrics, and compare predictions to existing experimental data sets available in the literature.

\label{sec:Results}

\subsection{Comparison to experiment}
\begin{figure}
    \centering
    \includegraphics[width=1\textwidth]{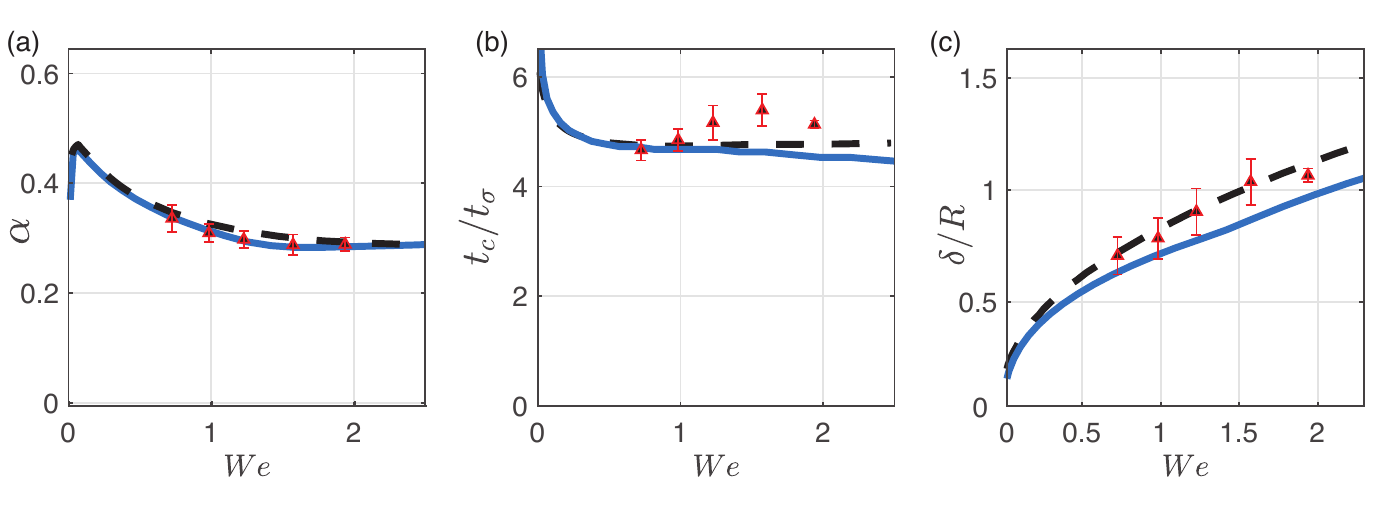}
    \caption{(a) Coefficient of restitution, (b) contact time, and (c) maximum penetration depth for a $R=0.35$ mm deionized water droplet rebounding from a bath of the same fluid as a function of $We$ (with $\Bo$ $= 0.017$ and $\Oh$ $= 0.006$). Error bars on experimental data points are quantified as the standard deviation of at least 5 experimental trials. Predictions of the quasi-potential model are shown as blue solid lines, DNS as black dashed lines. }
    \label{fig:summaries_withdata_W}
\end{figure}

\begin{figure}
    \centering
    \includegraphics[width=1\textwidth]{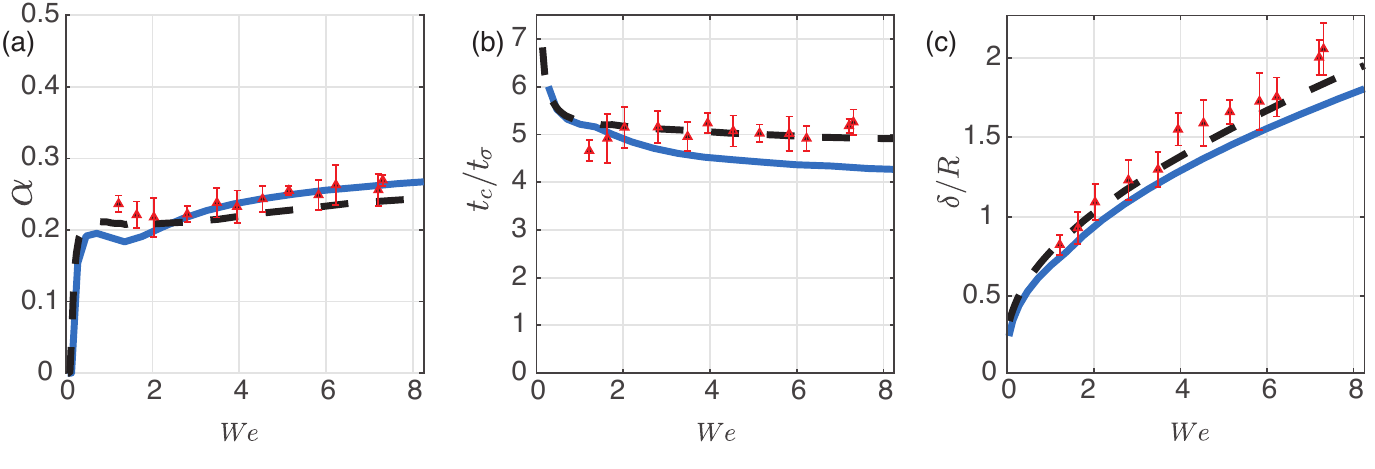}
    \caption{(a) Coefficient of restitution, (b) contact time, and (c) maximum penetration depth for a $R=0.35$ mm 5 cSt oil droplet rebounding from a bath of the same fluid as a function of $We$ (with $\Bo$ $= 0.056$ and $\Oh$ $= 0.058$). Error bars on experimental data points are quantified as the standard deviation of at least 5 experimental trials. Predicitions of the quasi-potential model are shown as blue solid lines, DNS as black dashed lines. }
    \label{fig:summaries_withdata_O}
\end{figure}
We first consider a single impact of a deionized water droplet with $R=0.35$ mm. A direct comparison of the trajectory results of the model, DNS, and the experiments is depicted in Figure \ref{fig:traj}(a). We find excellent agreement for the top, center of mass, and bottom trajectories between the experiment, quasi-potential model, and DNS. Additionally, we make a direct comparison between the quasi-potential model and the DNS for the prediction of the radius of contact in Figure \ref{fig:traj}(b). During impact, the quasi-potential model accurately predicts the instantaneous contact area, as well as the maximum contact area. The DNS, which accurately resolves the air film, shows that a finite region of contact is already developed prior to impact. At $t/t_{\sigma} > 3$, the two models deviate from each other, and this is most likely due to suction effects as the droplet pulls away from the surface. These effects cannot be predicted while neglecting the motion of the air, as in the current quasi-potential model, however do not appear consequential to the overall droplet trajectory. A comparison between the two models for the maximum width of the droplet is depicted in Figure \ref{fig:traj}(c). As is the case with the contact radius, the deformation of the droplet in the DNS occurs slightly before that of the quasi-potential model, as the pressure in the film is already building prior to contact. This leads to an overall phase shift of the oscillation between the models, although the maximum value for the deformation between both models remains in very good agreement.  We also compare the droplet and interface shape between the experiments, DNS, and linearized model can be seen in the panels of Figure \ref{fig:traj}(d). The impact is depicted for six different instants during contact. As the droplet deforms the interface, a capillary wave travels from the impact location to the north pole of the droplet. The collapse of this wave onto itself occurs just before the time of maximum deformation of the bath, and corresponds to the maximum deformation of the droplet. After this time, surface tension in the bath begins to act to restore the equilibrium, having redistributed some of the initial impact energy in the form of interfacial waves. The droplet remains mostly spherical as the bath relaxes, until contact is lost. During free flight, the droplet oscillates as an underdamped harmonic oscillator, dissipating additional energy through viscosity. Both the DNS and experiment show slightly stronger oscillations in the top of the droplet as compared to the model, to the point which the instantaneous slope at the top is occasionally close to zero. Overall, the DNS and quasi-potential model are in excellent agreement and predict the bath shape, droplet shape, and droplet trajectory with high accuracy for these parameters. \\
\hspace*{1em} As we begin to explore the larger parameter space, we consider three different output parameters for the rebounds, namely: coefficient of restitution ($\alpha$), contact time ($t_c$), and maximum surface deflection ($\delta$). As mentioned in section \ref{Section:ExpProc}, given the experimental difficulty to accurately determine the time of surface detachment of the droplet, contact time, $t_c$, is defined in the experiment as the interval between the two instances when the north pole of the droplet crosses level $z=2R$ and the coefficient of restitution, $\alpha$, is defined as minus the ratio of the vertical velocities at those times. For the model and DNS, we define the metrics in the same way, but when the center of mass of the droplet crosses $z=R$. This is chosen because a measure on the center of mass more accurately describes the total translational energy transfer from the droplet.  However, in comparing the results from the model and DNS using both top (measured at $z=2R$) and center of mass (measured at $z=R$), we found a typical difference of $2\%$ for $\alpha$ and $t_c/t_{\sigma}$ in the silicone oil experiments, and $5\%$ for the same parameters in the deionized water experiments. 

Figure \ref{fig:summaries_withdata_W} outlines a variation of impact $\We$ for a deionized water droplet onto a water bath. In this parameter sweep, the coefficient of restitution generally decreases as the $\We$ number is increased, eventually saturating at a value just below $0.3$, and remaining nearly independent of the $\We$ number thereafter. The contact time also decreases as the $\We$ number is increased, but remains relatively independent of the $\We$ number at an earlier value, consistent with results found for impact on a solid surface \citep{richard2002contact} and for previous experiments for impact on a deep pool \citep{zhao2011transition}. The maximum penetration depth increases monotonically with $\We$. We find good agreement between the model, DNS, and experiments with regard to the restitution coefficient, contact time, and maximum surface deflection for these experiments. Additionally, the quasi-potential model is able to predict $\alpha$ for the entire range of experiments. Both the model and DNS slightly underpredict the dimensionless contact time at intermediate $\We$, yet do agree with the experimental data for $\We$ $\leq 1$.  The quasi-potential model also underpredicts the penetration depth and contact time at moderate $\We$. Nonlinear effects associated with larger deformation have been neglected in this model, and are most likely the cause of this discrepancy.

Figure \ref{fig:summaries_withdata_O} depicts a $\We$ number variation using $5$ cSt silicone oil.  In non-dimensional terms, this represents an increase in both the $\Bo$ and $\Oh$ numbers.  Similar trends in $t_c/t_{\sigma}$ and $\delta$ are observed for the silicone oil as compared to the deionized water.  However, $\alpha$ tends to generally increase with $\We$ in this case, as opposed to water which showed the opposite trend. The quasi-potential model accurately predicts $\alpha$ for almost the full range of $\We$, with slight underprediction for $\We$ $<2$.  Similar to the water case, the quasi-potential model underpredicts $t_c/t_{\sigma}$ and $\delta$ at intermediate $\We$, with the DNS capturing these metrics more accurately. Although not verified experimentally, the model and DNS predict that droplets with very, very low $\We$ numbers cease to return to their original height at all.  Note that $\We$ $\leq 0.5$ is challenging to explore experimentally for these parameters, as pinch off of the droplets from the generator induces oscillations that need to dampen out prior to impact. The short free flight time and low viscosity of these extremely low $\We$ cases mean that there is still oscillation present at impact, which has been shown in prior work to influence rebound dynamics in related droplet impact problems \citep{biance2006elasticity,yun2018impact}.

\subsection{Inertio-capillary limit}
\begin{figure}
    \centering
    \includegraphics[width=1\textwidth]{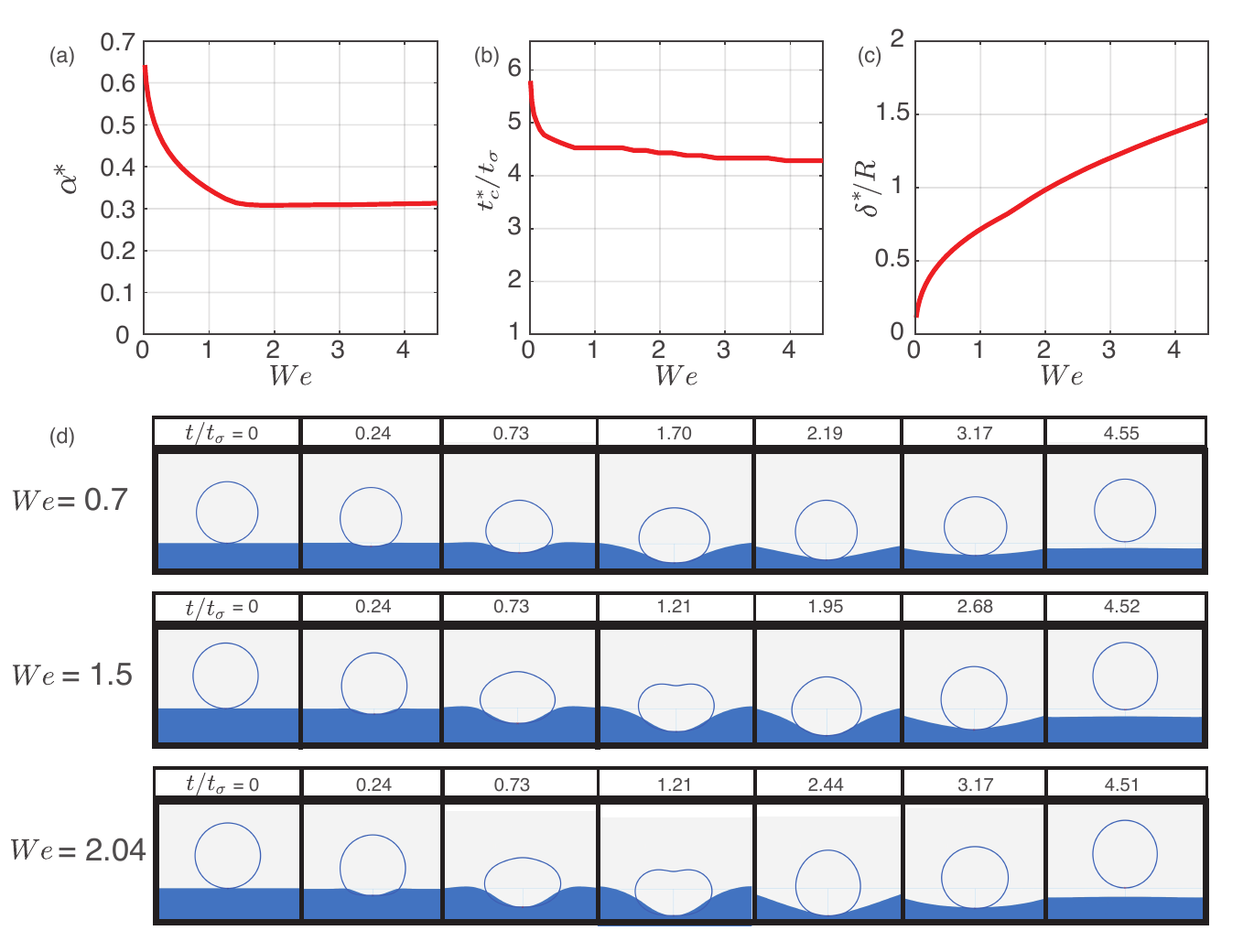}
    \caption{(a-c) Rebound parameters of a droplet in the inertio-capillary regime (as denoted by the asterisk). d) Droplet and bath shape in the pure inertio-capillary regime ($\Bo$ $= 0$, $\Oh$ $= 0$) for three different $We$.}
    \label{fig:ic_balance}
\end{figure}

The validated quasi-potential model can now be used to explore other sets of parameters.  In particular, based on the assumptions of the model, we expect the model to remain accurate (and even perhaps improve) for cases of even smaller $\Bo$ and $\Oh$ than achieved in experiments.  As a grounding point, we first turn our attention to the pure inertio-capillary limit, where both gravitational and viscous effects are ignored (i.e. $\Bo=0$ and $\Oh=0$).  This case reduces the number of dimensionless parameters that describe the physical problem to just one: the $\We$ number. The results in the inertio-capillary limit are presented in Figure \ref{fig:ic_balance}, along with the droplet and bath shape predictions for various $\We$. The penetration depth $\delta^*$ increases monotonically with increasing $\We$, as found for both the water and oil experiments. Additionally, the contact time $t_c^*$ decreases before becoming mostly independent of $\We$. However, in this limiting case, the coefficient of restitution $\alpha^*$ monotonically decreases with $\We$ and does not have a local maximum in the restitution coefficient, and droplets can even retain a majority of their impacting energy at sufficiently low $\We$. Furthermore, the coefficient of restitution then remains nearly independent of $\We$ above $\We$ $>1.75$, and is predicted to saturate to a value of approximately $\alpha_{s}^* = 0.31$.


\subsection{Influence of viscosity and gravity}
\begin{figure}
    \centering
    \includegraphics[width=0.975\textwidth]{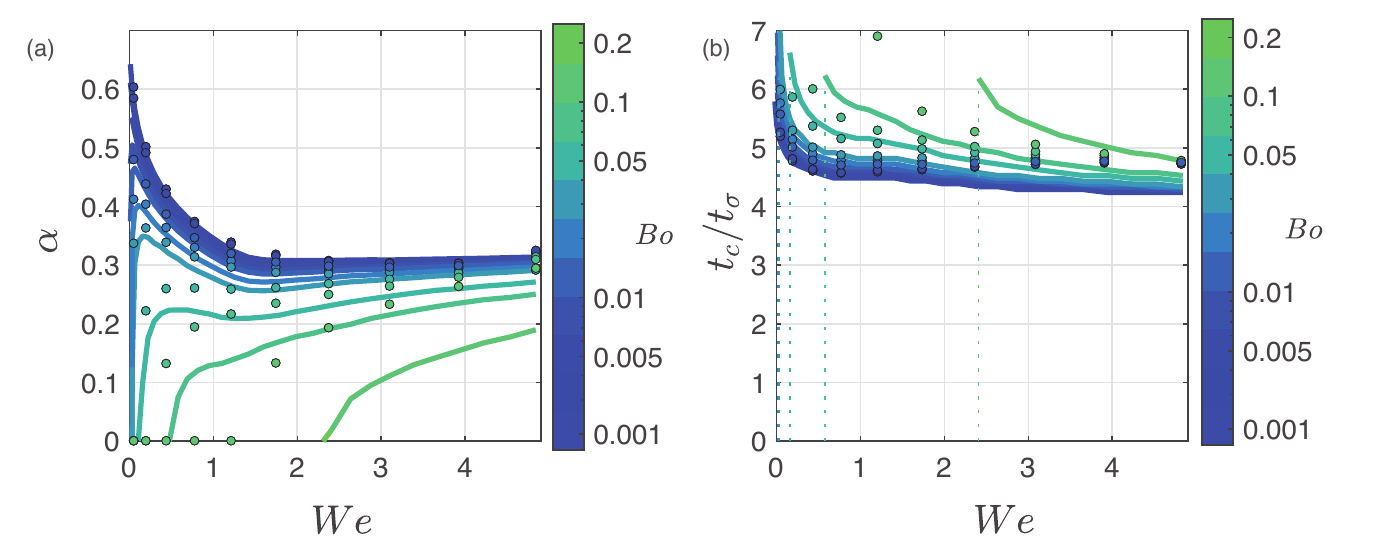}
    \caption{(a) Coefficient of restitution and (b) contact time as a function of the Bond number $\Bo$. Viscous effects are set to be finite but negligible, with $\Oh$ $=6\times10^{-4}$. Predicitions of the quasi-potential model are shown as solid lines, DNS as individual markers. The vertical dashed lines in panel (b) reference the critical $\We$ for each $\Bo$ below which droplets do not bounce.}
    \label{fig:sweepBo}
\end{figure}

\begin{figure}
    \centering
    \includegraphics[width=0.95
    \textwidth]{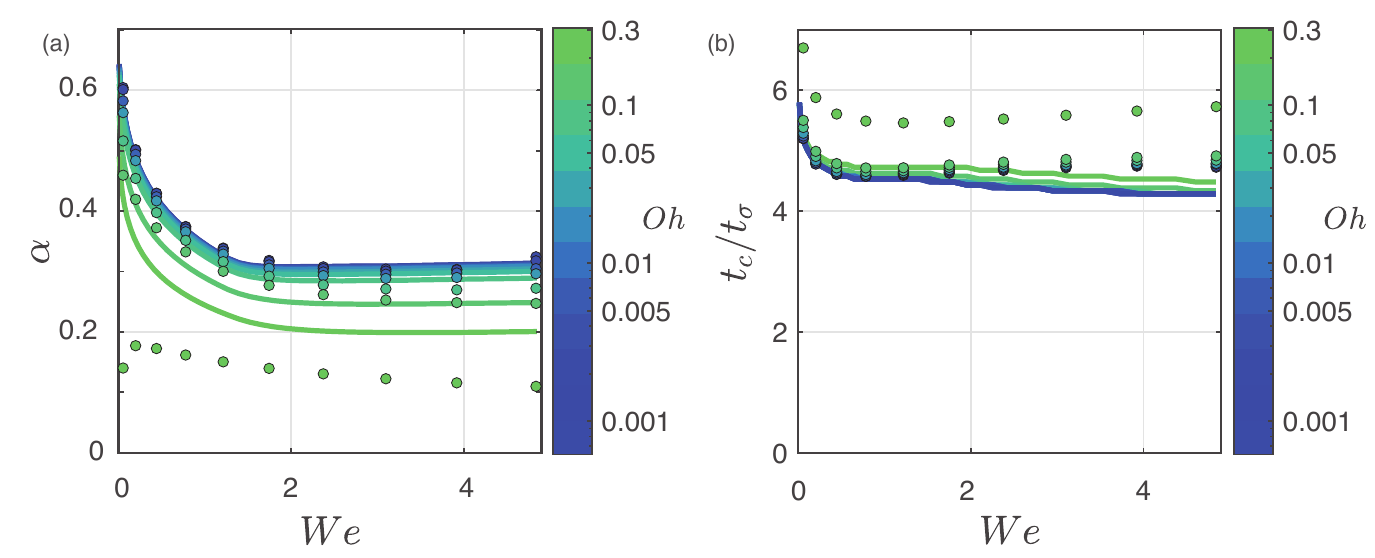}
    \caption{(a) Coefficient of restitution and (b) contact time as a function of the Ohnesorge number $\Oh$. Gravitational effects are set to be finite but negligible, with $\Bo$ $= 1 \times 10^{-3}$. Predicitions of the quasi-potential model are shown as solid lines, DNS as individual markers.}
    \label{fig:sweepOh}
\end{figure}

We then individually probe the parameter space by increasing either $\Bo$ or $\Oh$. These variations are presented in Figures \ref{fig:sweepBo} and \ref{fig:sweepOh}. The result of increasing $\Bo$, while keeping $\Oh$ constant (yet negligibly small), is depicted in Figure \ref{fig:sweepBo}.  At low values $\Bo$, the curves converge to the inertio-capillary limit as presented in the prior section.  For a given $\We$, the coefficient of restitution decreases monotonically with $\Bo$, until eventually ceasing to return to its original height.  At intermediate values of $\Bo$ the curves exhibits an interesting non-monotonic dependence on $\We$, with a local maxima at finite $\We$. Furthermore, the contact time of the droplets is predicted to increase with increasing $\Bo$. These qualitative trends are consistently reproduced by the DNS, with satisfactory quantitative agreement between the quasi-potential model and DNS for $\Bo \lesssim 0.1$.  For larger $\Bo$, and for the highest $\We$ cases explored, the model generally underpredicts the coefficient of restitution as compared to the DNS.
 
 Increasing $\Oh$ while keeping $\Bo$ negligibly small, is shown in Figure \ref{fig:sweepOh}, and also predicts a monotonic decrease in the restitution coefficient, with curves converging to the inertio-capillary limit for small $\Oh$. Unlike the $\Bo$ variation, the shape of the curve remains relatively unchanged. The non-dimensionalized contact time changes only marginally, even over an order of magnitude increase in $\Oh$. Very similar trends are predicted by DNS, with models diverging quantitatively beyond $\Oh \gtrsim 0.1$, consistent with a breakdown of the weakly viscous modeling assumptions.  For larger $\Oh$, the quasi-potential model overpredicts the coefficient of restitution and underpredicts the contact time, as compared to the DNS.
 

\subsection{Scaling analysis}
In this subsection we present scaling arguments to rationalize the dependence of the coefficient of restitution on $\Bo$ and $\Oh$ detailed in Figures \ref{fig:sweepBo}(a) and \ref{fig:sweepOh}(a), respectively.  As revealed in the prior section, at a fixed $\We$, the coefficient of restitution decreases monotonically from the inertio-capillary limit (Fig. \ref{fig:ic_balance}(a)) as either $\Bo$ or $\Oh$ is increased.  Due to the number of parameters involved, in order to proceed, we will assume that this additional energy transfer (or loss) due to weak gravitational (or viscous) effects is independent of the baseline energy transferred ($\Delta E^*$) in the inertio-capillary limit.  Mathematically, this assumption can be expressed in the form
\begin{equation}
    \alpha^2 = \frac{E_o - \Delta E^* - \Delta E_{g,\mu}}{E_o} = \alpha^{*2}-\frac{\Delta E_{g,\mu}}{E_o},
\end{equation}
or
\begin{equation}
    \alpha^{*2}-\alpha^2 = \frac{\Delta E_{g,\mu}}{E_o},\label{energy}
\end{equation}
where $E_o=\frac{2\pi}{3}\rho R^3 U^2$ is the initial droplet kinetic energy, and $\Delta E_{g,\mu}$ is the supplemental energy transferred or lost due to gravity or viscosity, respectively.  In what follows, we propose scalings for these energies.

\begin{figure}
    \centering
    \includegraphics[width=1
    \textwidth]{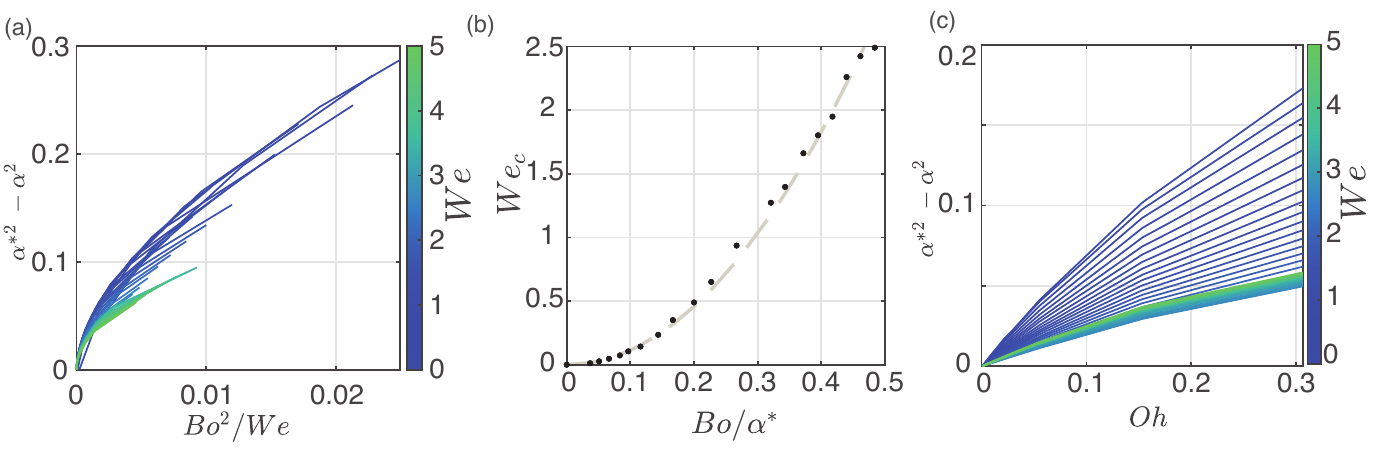}
    \caption{(a) Coefficient of restitution predictions from the quasi-potential model for $\Oh$ $=6\times10^{-4}$ (Fig. \ref{fig:sweepBo}(a)) re-plotted as informed by the scaling in equation (\ref{energyBo}). (b) Critical $\We$ as a function of $\Bo$; the points are predictions from the quasi-potential model, and the dotted curve is a parabolic fit ($\We_c = 11.43 \Bo^2/\alpha^{*2}$) consistent with the scaling in equation (\ref{criticalWe}).  (c) Coefficient of restitution predictions from the quasi-potential model for $\Bo$ $= 1 \times 10^{-3}$ (Fig. \ref{fig:sweepOh}(a))re-plotted as informed by the scaling in equation (\ref{energyOh}).}
    \label{fig:scaling3panel}
\end{figure}

Since gravity is a conservative force, increases in $\Bo$ must lead to an overall increase in the deformation (and subsequent oscillation) of the bath and droplet.  In the capillary-dominated regime, the gravity-induced deformation can be estimated to scale with $\rho R^3 g/\sigma$, corresponding to an additional surface energy scaling as $\Delta E_g \sim \rho^2 R^6 g^2 / \sigma$.  Normalizing this estimate by the incident kinetic energy, we find an additional fractional energy transfer to droplet-bath deformations that scales as
\begin{equation}
    \frac{\Delta E_g}{E_o} \sim \frac{\rho R^3 g^2}{\sigma U^2}=\frac{\Bo^2}{We}.\label{energyBo}
\end{equation}
Motivated by equations (\ref{energy}) and (\ref{energyBo}), we re-plot the data from Figure \ref{fig:sweepBo}(a) in \ref{fig:scaling3panel}(a), and find a satisfactory collapse.  In particular, the scaling in equation (\ref{energyBo}) correctly predicts that bounces at lower $\We$ will be more heavily penalized in terms of their coefficient of restitution.  This inequity rationalizes the observed non-monotonic behavior of $\alpha$ with $\We$ predicted for intermediate $\Bo$ (Figure \ref{fig:sweepBo}(a)).
Furthermore, equations (\ref{energy}) and (\ref{energyBo}) suggest a scaling for the ``critical'' Weber number $\We_c$ below which the droplet ceases to bounce (i.e. $\alpha^2<0$):
\begin{equation}
    We_c \sim \frac{Bo^2}{\alpha^{*2}}.\label{criticalWe}
\end{equation}
The data for the critical Weber number as predicted by the quasi-potential model is shown in Figure \ref{fig:scaling3panel}(b), and follows the proposed scaling in equation (\ref{criticalWe}).  We note that \cite{blanchette2016modeling} observed a parabolic scaling for the critical Weber number with $\Bo$ in prior work, also finding this threshold to be largely independent of $\Oh$.

Upon the inclusion of viscosity, there is viscous dissipation in the drop and bath that now occurs during contact.  The rate of viscous energy dissipation (per unit volume) can be estimated to scale as $\mu (\nabla \mathbf{u})^2\sim \mu U^2 / R^2$.  Assuming a characteristic fluid volume $R^3$, we find a viscous energy dissipation rate that scales like $ \mu U^2 R$.  As demonstrated in the prior sections, the contact time $t_c \sim t_\sigma$, and thus we may estimate the additional fractional energy loss during contact as
\begin{equation}
    \frac{\Delta E_\mu}{E_o} \sim \frac{\mu}{\sqrt{\sigma \rho R}}=Oh.\label{energyOh}
\end{equation}
Replotting the data from Figure \ref{fig:sweepOh}(a) in \ref{fig:scaling3panel}(c) shows that apart from the very lowest $We$ cases considered, the curves collapse to a single line, confirming the proposed scaling.  For all $\We$, the curves are approximately linear in $\Oh$ (consistent with equation (\ref{energyOh})) with the slope evidently depending on $\We$ for the smallest $We$ cases.  

In summary, our models predict an additional energy transfer (or loss) over the pure inertio-capillary limit when gravitational (or viscous) effects are introduced.  Our scaling suggests that gravity leads to additional deformations in the system, coming at an additional energetic cost.  Additionally, viscosity provides a mechanism for energy dissipation, occuring over the finite contact time of the droplet.  Computing the various energies directly in direct numerical simulations (such as in \cite{sanjay2022drop}) may provide additional insight to the remaining subtleties present in the data. 

\subsection{Comparison to prior literature data}

\begin{figure}
    \centering
    \includegraphics[width=0.95\textwidth]{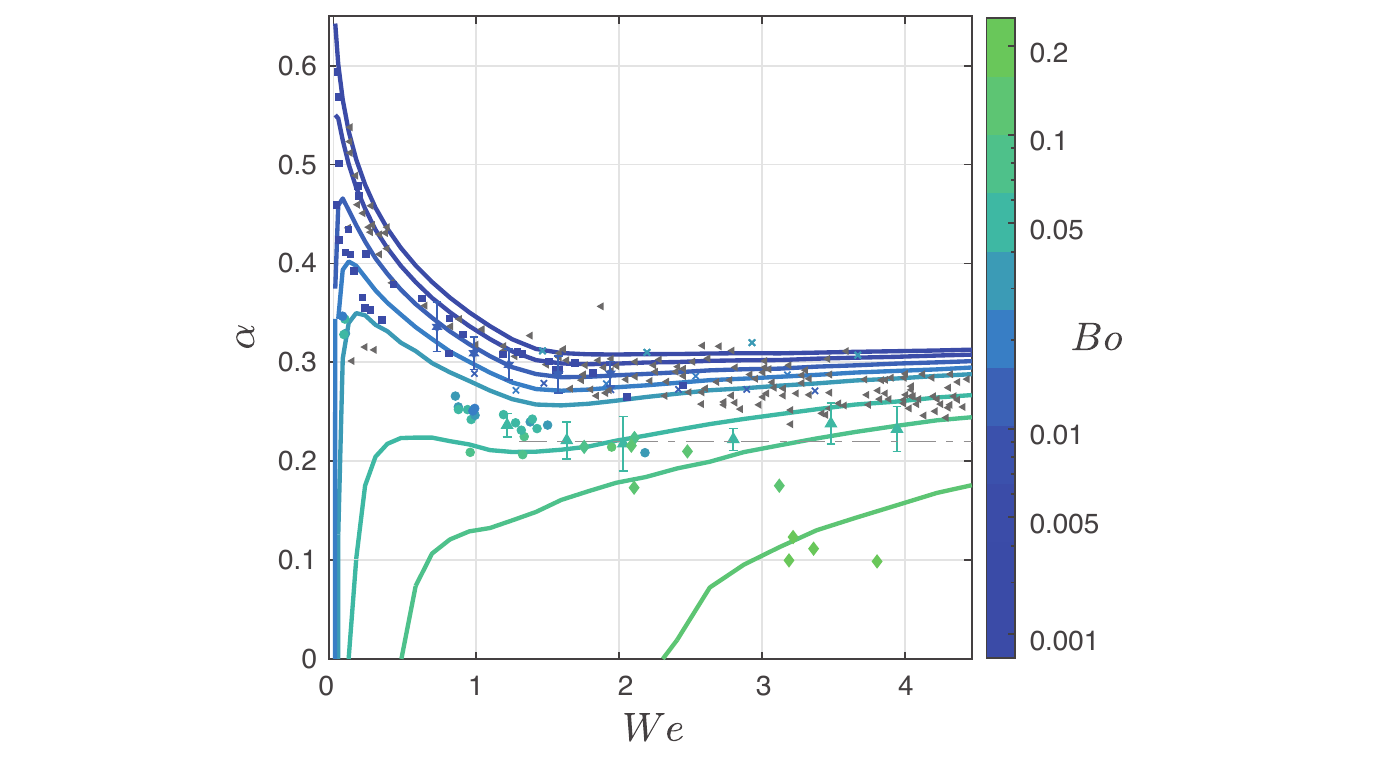}
    \caption{Comparison of model $\Bo$ predictions (Figure \ref{fig:sweepBo}(a)) to existing literature data where $We$ was reported. The grey markers represent data points where the exact value of $\Bo$ is unknown, with only ranges reported. The grey dot-dashed line represents the extrapolation of oblique impact data to normal impacts by \cite{jayaratne1964coalescence}. Data from the experiments completed in the present work are included with error bars. Studies included here are summarized in Table \ref{tab:litdata}}
    \label{fig:litdata}
\end{figure}

\hspace*{1em} In the works of \cite{jayaratne1964coalescence,bach2004coalescence,zhao2011transition,zou2011experimental,molavcek2013drops} there is a reported saturation in the energy transfer from the drop during rebound at modest $\We > 1$ and low $\Oh$, as measured by the coefficient of restitution.  The exact value of the saturation restitution coefficient does seem to vary however, from $0.2$ in \cite{molavcek2012quasi} for more viscous drops, to $0.3$ in \cite{bach2004coalescence,zhao2011transition}, to as low as $0.1$ \citep{zou2011experimental} for large $\Bo$ impact scenarios.  Remarkably, recent experiments on rebound of liquid metal droplets in viscous media also showed similar values of the coefficient of restitution with $\alpha=0.27$ \citep{mcguan2022dynamics}.  A similar saturation is also observed in our experimental results presented herein, with water droplets generally bouncing higher that the 5 cSt silicone oil droplets.  In Figure \ref{fig:litdata}, we overlay existing available experimental data for $\alpha$ from numerous sources and find that the prediction from the quasi-potential model
accurately captures much of this data.  The grey line represents the extrapolation of data from non-normal impacts by \citep{jayaratne1964coalescence} over the range of $\We$ reported in their work.  Data from the experiments completed in this work utilizing $5$ cSt silicone oil and deionized water are included with error bars.  The historical data generally indicates a decrease in $\alpha$ with $\Bo$, as captured by the present model.  Despite the relatively large variation in $\Oh$, the experimental data appear to match the results of the quasi-potential model quite well.

Furthermore, the existing experimental data (apart from a small number of outlying points) is well bounded by the inertio-capillary limit presented earlier.  This curve thus appears to define a universal upper bound on the coefficient of restitution for a droplet rebounding from a deep pool of the same fluid as a function of $\We$, regardless of any other parameters.

There is substantially less data available on contact times for low $\Oh$ impacts, but when reported they generally take values within the range of $4-6 t_{\sigma}$ \citep{zhao2011transition,molavcek2013drops,wu2020small}.

\begin{table}
\centering
\caption{Relevant publications and ranges of droplet parameters for experiments presented in Figure \ref{fig:litdata}.}
\begin{tabular}{l|c|c|c}
\hline
\textbf{Publication}       
& \textbf{$Bo$} 
& \textbf{$Oh$}
& \textbf{Symbol}
\\
\cite{bach2004coalescence}
& 
$5 \times 10^{-5} - 2 \times 10^{-4}$
& 
$0.31 - 0.69$
&
$\blacksquare$
\\   
\cite{zhao2011transition}
& $> 0.005$
& $> 0.008$
&$\vartriangleleft$
\\
\cite{zou2011experimental}
& 
$0.10-0.48$
& 
$0.0026-0.0039$
& $\diamond$
\\   
\cite{molavcek2013drops}
& 
$0.027-0.2$
& 
$0.16-0.27$
& $\circ$
\\
\cite{wu2020small}
& 
$0.016-0.04$
& 
$0.005-0.006$
& $\times$
\\   
This work
& 
$0.017-0.056$
& 
$0.006-0.058$
& $\vartriangle$
\\   
\end{tabular}
\label{tab:litdata}
\end{table}

\section{Discussion}
In this work we have used a combination of custom experiments, state-of-the-art direct numerical simulation techniques, and a new quasi-potential model to study the bouncing of millimetric droplets on a deep bath of the same fluid in the inertio-capillary regime. Weakly viscous models for the bath and droplet interfaces are coupled to one another through the use of a simplified kinematic matching condition, and allow us to make accurate predictions for the droplet trajectory and time-dependent droplet and bath shapes. Furthermore, the quasi-potential model is relatively efficient to compute and uses only standard off-the-shelf algorithms, resolving multiple bounces in less than 5 minutes on a standard desktop computer.  Both the quasi-potential model and DNS are demonstrated to accurately predict the coefficient of restitution, contact time, and maximum surface deflection, validated by direction comparison to experiments with two different working fluids. 

Starting from the inertio-capillary limit (where both gravitational and viscous are negligible) in the model, as we increase $Bo$, we see a decrease in the coefficient of restitution and an increase in the contact time. Additionally, a local maximum develops in $\alpha$ at low $\We$ as $Bo$ increases. As $Bo$ increases further, droplets cease to return to their original height. Furthermore, as $Oh$ is increased away from the inertio-capillary limit, a simpler, monotonic decrease in $\alpha$ is observed, with the contact time remaining almost unchanged for the much of the $Oh$ range explored in this work.  By further comparison with DNS, the complete model is shown to hold in the limit of small $Bo$ and $Oh$, and up to intermediate $We$, provided that the influence of the intervening gas layer that inhibits coalescence on the overall dynamics is minimal. These trends can be rationalized using simple scaling arguments, with gravity resulting in additional droplet-bath deformations (and thus energy transfer), and viscosity providing a mechanism for energy dissipation in the fluid during contact. Additionally, the model can be used to connect much of the existing experimental data on this particular topic.  In particular, the inertio-capillary limit appears to define an upper bound on the possible coefficient of restitution for droplet-bath impact, with the value depending only on the Weber number, and saturating to a near constant value at intermediate $We$.

The related problem of a droplet impactor rebounding off a solid surface has been considered in numerous previous works \citep{anders1993velocity,richard2000bouncing,richard2002contact,gilet2012droplets}. The dependence of the coefficient of restitution on the Weber number has previously been reported \citep{biance2006elasticity,aussillous2006properties,gilet2012droplets}, and the trend observed in these studies is similar to what is found in this work for low $\We$ and low $\Bo$, however the typical values of restitution coefficients in these studies are significantly larger. This is likely due to the fact that a large portion of the initial droplet energy in the present case is carried away by surface waves excited in the fluid bath. Our general findings also have many similarities with the investigation of \citet{galeano2021capillary}, in which non-wetting spheres impact and rebound from a water bath.  In particular, the general trends for maximum penetration depths and contact times are consistent with the present work.  However, spheres with density most similar to that of water show a consistent monotonic increase in the coefficient of restitution with increasing $We$ rather than saturating to a near constant value for the case of droplet-bath rebound.  Furthermore, at intermediate $\We$, coefficients of restitution can take values as high as $\alpha \approx 0.5$, distinctly greater than otherwise equivalent droplet-bath rebounds considered in the present work.  Evidently, the nature of the impactor and substrate influences the subtle energy transfer mechanisms across these different capillary rebound problems.

The model presented here is highly versatile, with only a single embodiment thereof considered here in terms of target canonical physical scenario.  Future work will consider the effect of relative surface tension and viscosity, where the droplets are composed of a different fluid than the bath. Also, in the present work we have specifically selected the size of the bath to be much greater than that of the droplet radius to eliminate any possible wave reflection and interaction effects. In \cite{zou2013phenomena}, experimental results indicate that reducing the size of the bath (such that the impacting wave on the bath has time to travel to the edge and return to the impact point) can increase the coefficient of restitution with the droplet recovering energy initially lost to waves.  Furthermore, the effects of incident droplet or bath deformations could be readily studied, and has been shown to influence bouncing behavior in similar systems \citep{biance2006elasticity,yun2018impact}.

There are a number of possible additions to the existing model that could expand its reach to other related problems.  For instance, the model for the droplet deformation can be extended into a regime where the dynamics of the gas layer does matter, and the gas layer dynamics coupled to the droplet deformation through the use of lubrication equations. The model can also be adapted to non-axisymmetric domains, or to droplet impacts at varying angles of incidence. However, in these cases the full kinematic match should be utilized, as the shape of the pressure distribution would likely change at each time step.  Moreover, numerous authors have studied the variety of phenomena that occur when a droplet impacts another droplet \citep{qian1997regimes,tang2012bouncing}. Droplet-droplet collisions are of extreme importance in combustion science \citep{jiang1992experimental} and cloud formation \citep{grabowski2013growth}, for instance, the general effect of cloud turbulence acts to increase droplet-droplet interaction, and droplet impact and coalescence is postulated as the primary mechanism by which warm rain forms \citep{grabowski2013growth}. With such motivation in mind, the present model could be readily extended to cases where equal and unequal sized droplets impact and rebound from one another. Overall, the quasi-potential model developed in this work has the potential to continue to inspire and inform the rich subject of capillary rebounds.

\vspace{5mm}
{\bf Supplementary material.} Supplementary material (experimental, model and simulation animations) are made available to the interested reader.

\vspace{5mm}
{\bf Acknowledgements.} The authors would like to acknowledge Carlos Galeano-Rios for his support and guidance throughout the project, Ajay Harishankar Kumar for helpful discussions, Fran{\c{c}}ois Blanchette for sharing code associated with prior work, and Jan Mol{\'a}{\v{c}}ek for sharing experimental data from prior work. The Imperial College London Research Computing Service and the Scientific Computing Research Technology Platform at Warwick have both supported this work with computational resources. All authors would like to thank the referees for their constructive suggestions.

\vspace{5mm}
{\bf Funding.} The authors gratefully acknowledge the financial support of the National Science Foundation (NSF CBET-2123371) and the Engineering and Physical Sciences Research Council (EPSRC EP/W016036/1). RC also thanks the London Mathematical Society for early stage funding through a Scheme 4 Research in Pairs Grant (Ref. 41813).

\vspace{5mm}
{\bf Code accessibility.} Codes and associated documentation can be found at \url{https://github.com/harrislab-brown/BouncingDroplets} for the mathematical model implementation (MATLAB) and \url{https://github.com/rcsc-group/BouncingDroplets} for the direct numerical simulation framework (Basilisk C) used as part of this work. 

\vspace{5mm}
{\bf Declaration of interests.} The authors report no conflict of interest.

\newpage

\appendix 

\section{Influence of ambient gas properties}

\begin{figure}
    \centering
    \includegraphics[width=01\textwidth]{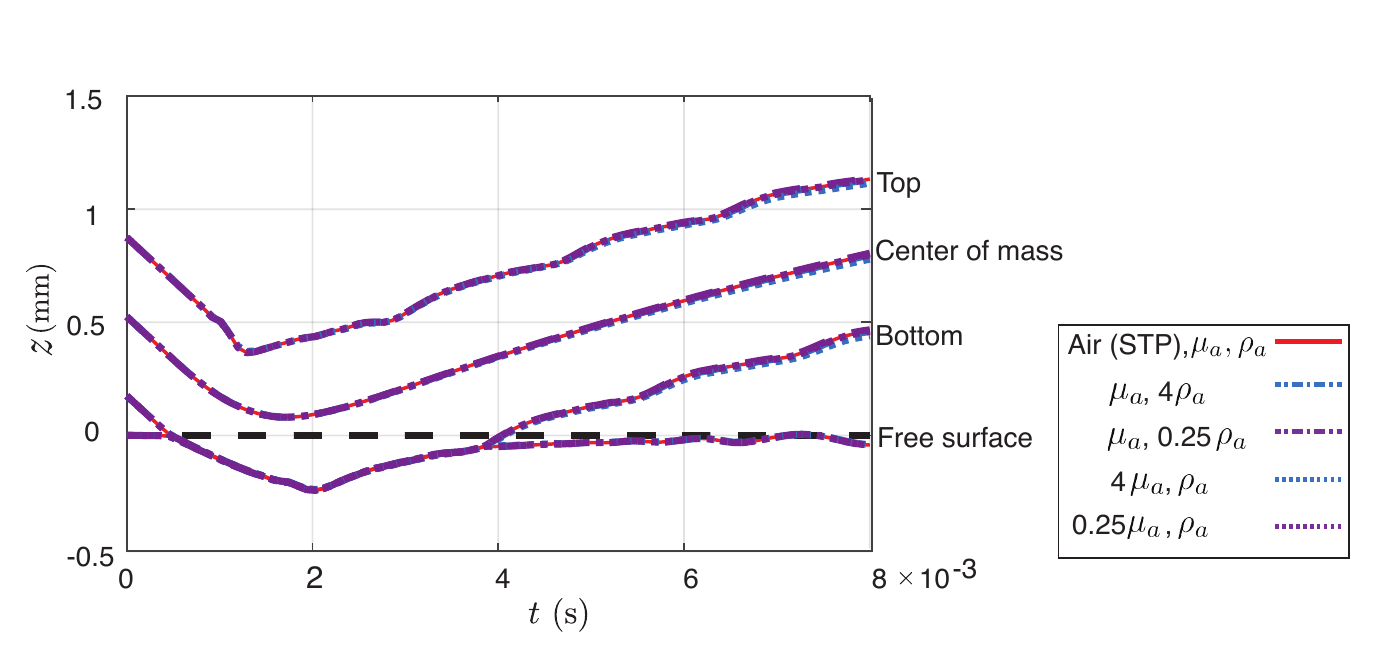}
    \caption{DNS predictions of the trajectory of a deionized water droplet with $R=0.35$ mm in air at $\We$ $=0.7$ (with $\Bo$ $=0.017$, $\Oh$ $=0.006$). The ambient gas density and viscosity are increased and decreased independently by a factor of 4. These simulations are compared to the case with the reference case of air properties at standard temperature and pressure (STP).}
    \label{fig:air_props}
\end{figure}

\hspace*{1em} In order to verify our assumption that the flow within the air layer is negligible to the droplet and bath dynamics in our parameter regime of interest, we run DNS simulations where the ambient gas density and viscosity are varied. First, the gas viscosity is held fixed for air at $21^{\circ}$C and 1 atm and the density is increased by a factor of four, and then decreased by a factor of four, respectively. Then the variation process is repeated for the gas viscosity, with density held fixed. The results of these simulations are presented in Figure \ref{fig:air_props} and are nearly indistinguishable, particularly during contact.

\section{Influence of pressure shape function}
\label{app:pressureShape}

\begin{figure}
    \centering
    \includegraphics[width=1\textwidth]{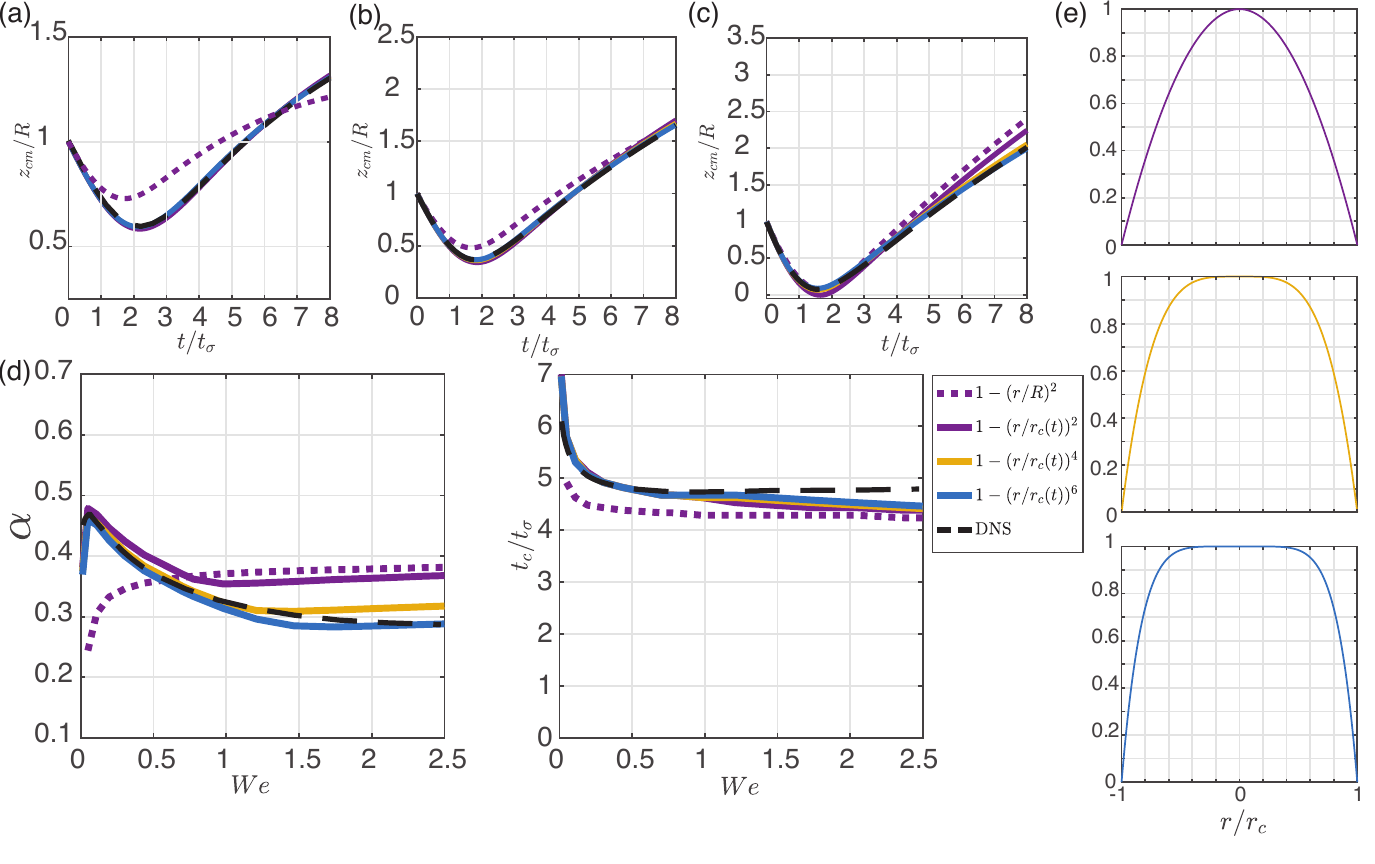}
    \caption{Quasi-potential model and DNS predictions for a deionized water droplet with $R=0.35$ mm corresponding to $\Bo$ $=0.017$ and $\Oh$ $=0.006$. Droplet center of mass trajectories for (a) $\We$ $=0.07$, (b) $\We$ $=0.36$, and (c) $\We$ $=1.36$. (d) Coefficient of restitution $\alpha$ and contact time $t_c/t_{\sigma}$ as a function of $We$. (e) Plots of the pressure shape functions $H_r(r/r_c)$ tested in this figure, shown for reference.}
    \label{fig:pressuredrop}
\end{figure}

 \hspace*{1em} Additionally, we tested several pressure shape functions $H_r$ to assess the relative influence of this choice. In the simulations show in Figure \ref{fig:pressuredrop} a parabola, 4th-order, and 6th-order polynomial with time varying contact areas set as described in the modeling section are compared as well as to a parabola with fixed contact radius $r_c(t) = R$. These results all correspond to $\Bo$ = $0.017$ and $\Oh$ = $0.006$ impacts. The parabola with the fixed contact area performs the most poorly, especially at lower impact $\We$. At higher impact $\We$, the fixed radius $r_c(t)=R$ parabolic distribution prediction becomes similar to the time evolving parabolic case. For the simulations presented in this work, the contact radius is defined to have a maximum value of $R$, as the projection of the pressure distribution onto the undisturbed spherical surface is no longer well defined for $r_c > R$. The value for $r_c(t)$ quickly saturates to $R$ at higher impact $\We$, and the agreement between the constant contact radius and the contact radius model used in the present work improves.  Overall, inclusion of a time-varying contact radius appears necessary to capture the correct trends in $\alpha$ and $t_c/t_{\sigma}$ over the range of $\We$ presented in this work. In addition, as the order of the polynomial increases, the shape of the pressure function more closely resembles a top hat, and the predictions of the model improves as compared to the corresponding DNS. The 6th-order polynomial was ultimately chosen for the present work, as higher polynomials (corresponding to broader flat regions) converged more slowly with only marginal changes in the quantitative predictions.

\newpage
\bibliographystyle{jfm}
\bibliography{Biblio}

\end{document}